\documentclass[preprints,article,accept,moreauthors,pdftex]{Definitions/mdpi} 

\firstpage{1} 
\pubvolume{1}
\issuenum{1}
\articlenumber{0}
\pubyear{2021}
\copyrightyear{2020}
\datereceived{} 
\dateaccepted{} 
\datepublished{} 
\hreflink{https://doi.org/} 

\newcommand{\be}{\begin{equation}}
\newcommand{\ee}{\end{equation}}
\newcommand{\bea}{\begin{eqnarray}}
\newcommand{\eea}{\end{eqnarray}}
\newcommand{\doublet}[2]{ \left( \begin{array}{c}#1 \\ #2 \end{array}\right) }
\Title{CP-Violation in the 3-Higgs Doublet Model: CP-Asymmetries from Charged Higgs Bosons and Electric Dipole Moments}

\Author{ Andrew G. Akeroyd$^{1}$, Heather E. Logan $^{2}$, Stefano Moretti $^{1,3}$, Diana Rojas-Ciofalo $^{4}$, Tetsuo Shindou $^{5}$,  and Muyuan Song$^{1,\dagger}$*}

\address{$^{1}$ \quad School of Physics and Astronomy,
University of Southampton, Highfield,
Southampton SO17 1BJ, United Kingdom \\
$^{2}$ \quad Ottawa-Carleton Institute for Physics, Carleton University, Ottawa, Ontario K1S 5B6, Canada \\
$^{3}$ \quad Particle Physics Department, Rutherford Appleton Laboratory, Chilton, Didcot, Oxon OX11 0QX,
United Kingdom \\ 
$^{4}$ \quad National Centre for Nuclear Research, Pasteura 7, 02-093 Warsaw, Poland \\
$^{5}$ \quad Division of Liberal-Arts, Kogakuin University,
2665-1 Nakano-machi, Hachioji, Tokyo, 192-0015, Japan}

\firstnote{Current address: Center for High Energy Physics, Peking University, Beijing 100871, China} 

\abstract{We propose a CP-violating source in the charged Higgs sector of a 3-Higgs Doublet Model (3HDM) that will contribute to three CP-asymmetry observables for low energy processes. In the context of such a 3HDM, a measurement of one of these specific asymmetry observables (in $\bar{B} \to X_{s+d} \gamma$ events) could attain up to 2.5\% or more, which would lead to a 5$\sigma$ signal for physics Beyond Beyond the Standard Model (BSM)  from the full BELLE II data set. In addition to these CP-violating observables, 
the current Electron Dipole Moments (EDMs) measurements constrain the available parameter region of the CP-violating 3HDM. In this work, we perform a thorough parameter space analysis of this BSM scenario to find the regions that satisfy the neutron and electron  EDMs (nEDM and eEDM, respectively), $\bar{B} \to X_s \gamma$ and direct collider searches for charged Higgs bosons as well as theoretical bounds.}


\begin{document}
\section{Introduction}

Charge/Parity (CP)-violation is a fascinating property to study in the realm of phenomenology in high energy physics, as it could point to new phenomena Beyond the Standard Model (BSM).    
In the SM, there is only one CP-violating source which comes from the Cabibbo-Kobayashi-Maskawa (CKM) quark mixing matrix, but this is not enough to obtain the matter-antimatter asymmetry of the universe~\cite{Sakharov:1967dj}.  
From the experimental measurements of several low energy processes, like the $\bar{B} \to X_s \gamma$ decay \cite{HFLAV:2019otj}, neutron Electric Dipole Moment (EDM) \cite{nEDM:2020crw} and electron one \cite{ACME:2018yjb}, we know there is a strict upper limit on the size of CP-violating effects from BSM scenarios. 
In particular, for the decay $\bar{B} \to X_s \gamma$, the Branching Ratio (BR) measured by experiment  is in excellent agreement with the prediction from the SM, providing a strong constraint on Multi-Higgs Doublet Models (MHDMs) charged sectors.  
For example, in the well known Type-II 2HDM, the authors of~\cite{Misiak2020} get a lower limit of 800 GeV on the $H^{\pm}$ mass. 
Moreover, 
{because of the Glashow-Iliopoulos-Maiani (GIM) cancellation and the CKM suppression, the SM contribution to CP-asymmetries in Flavour Changing Neutral Current (FCNC) processes is minimal.}
Presently, the estimation of the direct CP-asymmetry in $\bar{B} \to X_s \gamma$ for the SM is below 1\% of magnitude, thus lower than the experimental sensitivity~\cite{Kagan:1998bh}. 
Conversely, the theory prediction can significantly be enhanced when adding complex CP-violating phases in an extended scalar sector, which is an excellent motivation to investigate CP-violation effects in BSM frameworks.

The discovery in 2012 of a Higgs boson with a mass of roughly 125 GeV at ATLAS and CMS completed the experimental success of the SM \cite{ATLAS:2012yve,CMS:2012qbp}.  
Nevertheless, finding an extended Higgs sector remains a plausible possibility.
For this, it is essential to measure with high precision the Higgs BRs, from where one can explore the presence of additional doublet  or higher multiplet representations (e.g., additional singly and doubly charged ($H^{\pm}$ and $ H^{\pm\pm}$, respectively) or neutral (e.g., $H$ (CP-even) or $A$ (CP-odd)) Higgs bosons.  
The most studied MHDM in the literature is the 2HDM. In general, for this model, the non-observation of a charged scalar $H^{\pm}$ has already excluded a considerable parameter space corner in the $\tan\beta$ vs $M_{H^{\pm}}$  plane (for reviews, see \cite{Branco:2011iw,Akeroyd:2016ymd}), where
$\tan\beta = v_2/ v_1$, where $v_{1,2}$ are the two Higgs doublets Vacuum Expectation Values (VEVs).  

In the case of the 3HDM, the phenomenology of the ensuing two charged Higgs boson states depends on four free parameters ($\tan\beta, \tan\gamma, \theta, \delta$), which come from the unitary rotation matrix transforming the Higgs states from the Electro-Wea k(EW)  to mass physical eigenbasis~\cite{Grossman:1994jb,Akeroyd:1994ga,Akeroyd:2016ssd,Akeroyd:2018axd,Akeroyd:2019mvt}. 
An essential property of the 3HDM is a CP-violating phase $\delta$, which cannot be rotated away by rephasing the charged Higgs boson mixing parameters out of the initial four.
Due to the larger number of parameters and this new source of the CP-violation, the phenomenology involving the charged Higgs bosons is very different from the one in the 2HDM. 
In our work, we start with studying three CP-asymmetry observables relevant to $\bar{B} \to X_s \gamma$ decays, which involve the two charged scalars of the 3HDM. 
Here, we study the available parameter space for a clear potential discovery signal that could be seen at the BELLE-II experiment. 
Then, we analyse the EDM constraints (from neutron and electron) of two charged Higgs states coming from such a single CP-violation phase in the charged scalar sector. 

This paper is organised as follows. In Section \ref{section2}, we outline the 3HDM structure. We display the scalar potential and then discuss the charged Higgs spectrum and the CP-violating phases.
Then the processes showing the aforementioned CP-asymmetry observables are discussed while neutron/electron EDMs are illustrated in detail. In Section \ref{section3}, we show several results based on the calculation of CP-asymmetry and EDM constraints within the 3HDM the charged Higgs sector. Our overarching discussion of both asymmetry and EDM results is given in Section \ref{section4}.
Finally, Section \ref{section5} presents our conclusions. We then show the explicit coefficients for the different combinations of the charged Higgs Yukawa couplings in Appendix~\ref{appendix:char}. The contribution to $\bar{B} \to X_s \gamma$ decays from charged Higgs states is listed in Appendix~\ref{appendix:a}.  In Appendix~\ref{appendix:b}, we finally present the collider, perturbativity and top quark width constraints we used for our analysis.  
(This review is based on a combination of results from Refs.~\cite{Akeroyd2020} and \cite{Logan:2020mdz}.)

\section{Model and Method}\label{section2}
In this section, we describe the 3HDM and our approach to compute relevant observables in it.
\subsection{The Higgs Structure of the 3HDM}
In the 3HDM, there exist three $SU(2)_L $ scalar doublets:
\bea
 \Phi_i = \doublet{\phi^+_i}{(v_i + \phi^{0,r}_i + i\phi^{0,i}_i)/\sqrt{2}},\qquad i = 1,2,3.
\eea
The VEVs $v_i$ obey the sum rule $v^2_1 + v^2_2 + v^2_3 = v^2_{\text{SM}}$. 
In our model, there is Natural Flavour Conservation (NFC) (i.e., a single doublet only couples to a single type of fermion) to prevent FCNCs \cite{Glashow:1976nt,Paschos:1976ay}, which is imposed by three soft $Z_2$ symmetries.
The different assignments of the fermions interacting with the (pseudo)scalar Higgs states are prescribed by the Yukawa structure of five different ``Types'', as seen in Tab.~\ref{tab:NFC_list}.

\begin{specialtable}[H] 
	\begin{center}
	\caption{The five types of 3HDM under NFC condition.  This table implies that each Higgs doublet is responsible for a specific type of fermion mass generation. The symbol $u(d)$ means  up(down)-type quark while $\ell$ means a (charged)-type lepton.\label{tab:NFC_list}}
\begin{tabular}{cccc}
\toprule
{Model}	& {$\Phi_1$}	& {$\Phi_2$} & {$\Phi_3$}\\
\midrule
Type-I & -- & $u, d, \ell$ & -- \\
Type-II & $d, \ell$ & $u$ & -- \\
Type-X or Lepton-specific & $\ell$ & $u,d$ & -- \\
Type-Y or Flipped & $d$ & $u, \ell$ & -- \\
Type-Z or Democratic & $d$ & $u$ & $\ell$ \\
\bottomrule
\end{tabular}
	\end{center}
\end{specialtable}

The most general  scalar potential invariant under SU(2)$_{L}\times$U(1)$_{Y}$ with imposed NFC contain 18 free parameters and is given by:
\bea
        V &=& m^2_{11}\Phi^\dagger_1\Phi_1 + m^2_{22}\Phi^\dagger_2\Phi_2 + m^2_{33}\Phi^\dagger_3\Phi_3 \nonumber\\ &-& [m^2_{12}\Phi^\dagger_1\Phi_2 + m^2_{13}\Phi^\dagger_1\Phi_3+ m^2_{23}\Phi^\dagger_2\Phi_3 + \text{h.c.}] \nonumber\\
         &+& \frac{1}{2}\lambda_{1}(\Phi^\dagger_1\Phi_1)^2 + \frac{1}{2}\lambda_{2}(\Phi^\dagger_2\Phi_2)^2 + \frac{1}{2}\lambda_{3}(\Phi^\dagger_3\Phi_3)^2 \nonumber\\
         &+& \lambda_{12}(\Phi^\dagger_1\Phi_1)(\Phi^\dagger_2\Phi_2) + \lambda_{13}(\Phi^\dagger_1\Phi_1)(\Phi^\dagger_3\Phi_3)+ \lambda_{23}(\Phi^\dagger_2\Phi_2)(\Phi^\dagger_3\Phi_3) \nonumber\\
         &+& \lambda'_{12}(\Phi^\dagger_1\Phi_2)(\Phi^\dagger_2\Phi_1) + \lambda'_{13}(\Phi^\dagger_1\Phi_3)(\Phi^\dagger_3\Phi_1) + \lambda'_{23}(\Phi^\dagger_2\Phi_3)(\Phi^\dagger_3\Phi_2) \nonumber\\
         &+& \frac{1}{2}[\lambda^{\prime\prime}_{12}(\Phi^\dagger_1\Phi_2)^2 + \lambda^{\prime\prime}_{13}(\Phi^\dagger_1\Phi_3)^2 + \lambda^{\prime\prime}_{23}(\Phi^\dagger_2\Phi_3)^2 + \text{h.c.}].
\eea
In the above formula, the $Z_2$ symmetries are broken softly by the $m^2_{ij}$ terms. This potential has six complex CP-violating phase parameters: three are the soft-breaking masses, $m^2_{12}$, $m^2_{13}$ and $ m^2_{23}$, and the other three  are the quartic couplings, $\lambda^{\prime\prime}_{12}$, $\lambda^{\prime\prime}_{13}$ and $\lambda^{\prime\prime}_{23}$. In the case of phase rotating the three fields ($\Phi_1$, $\Phi_2$ and $\Phi_3$), two of the six parameters can be eliminated\footnote{Based on the U(1)$_Y$ hypercharge symmetry, the phase rotation of all three scalar fields does not change the potential. We pick $v_3$ to be real and positive, which leads to a meaningful condition since only relative phases are physically accessible.} leaving four CP-violating phases as physical. Furthermore, we do not remove the imaginary components of two of the complex parameters within the total of six, instead, we take all three VEVs to be positive and real, which will not lead to any loss of generality.  We then manage the imaginary components of $m_{13}^2$ and $m_{23}^2$ as follows~\cite{Cree2011}: 
\begin{subequations} 
\bea \label{EqAr:realvevs}
 {\text{Im}}(m^2_{13}) &=& -\frac{v_2}{v_3}{\text{Im}}(m^2_{12}) + \frac{v_1 v_2^2}{2 v_3}{\text{Im}}(\lambda^{\prime\prime}_{12}) + \frac{v_1v_3}{2}{\text{Im}}(\lambda^{\prime\prime}_{13}),\\
 {\text{Im}}(m^2_{23}) &=& \frac{v_1}{v_3}{\text{Im}}(m^2_{12}) - \frac{v_1^2v_2}{2v_3}{\text{Im}}(\lambda^{\prime\prime}_{12}) + \frac{v_2v_3}{2}{\text{Im}}(\lambda^{\prime\prime}_{23}).
\eea
\end{subequations}

The Higgs sector CP-violation sources manifest themselves from the mixing between the two CP-odd and  three CP-even neutral Higgs states as well as the two charged Higgs states and come from the remaining four independent complex phases.  
In this paper, we focus on a constrained version of the 3HDM, in the sense that there will be no CP-violation sources in the neutral Higgs sector. The following three relations are imposed to achieve this condition: 
\begin{subequations} \label{EqAr:CPsplit}
   \bea 	
	{\text{Im}}(\lambda^{\prime\prime}_{13}) &=& -\frac{v^2_2}{v_3^2}{\text{Im}}(\lambda^{\prime\prime}_{12}), \\
	{\text{Im}}(\lambda^{\prime\prime}_{23}) &=& \frac{v^2_1}{v_3^2}{\text{Im}}(\lambda^{\prime\prime}_{12}), \\
	{\text{Im}}(m^2_{12}) &=& v_1v_2{\text{Im}}(\lambda^{\prime\prime}_{12}).
   \eea
\end{subequations}
The above relations leave only one independent CP-violating parameter and it can be taken as {\rm Im}($\lambda_{12}^{\prime\prime})$.  This remaining CP-violating phase will occur in the charged Higgs mass matrix and is responsible for CP-violation stemming from the couplings of the charged Higgs mass eigenstates.

Finally, the potential could possibly be minimized and three real parameters can be eliminated in favour of the (real) VEVs~\cite{Cree2011}:
\begin{subequations}
\bea
 m^2_{11} &=& \frac{v_2}{v_1}{\text{Re}}(m^2_{12}) + \frac{v_3}{v_1}{\text{Re}}(m^2_{13}) - \frac{v_1^2}{2}\lambda_1 \nonumber \\
  &&- \frac{v_2^2}{2}[\lambda_{12} + \lambda'_{12} + {\text{Re}}(\lambda^{\prime\prime}_{12})] 
  - \frac{v_3^2}{2}[\lambda_{13} + \lambda'_{13} + {\text{Re}}(\lambda^{\prime\prime}_{13})], \\
 m^2_{22} &=& \frac{v_1}{v_2}{\text{Re}}(m^2_{12}) + \frac{v_3}{v_2}{\text{Re}}(m^2_{23}) - \frac{v_2^2}{2}\lambda_2 \nonumber \\
 &&- \frac{v_1^2}{2}[\lambda_{12} + \lambda'_{12} + {\text{Re}}(\lambda^{\prime\prime}_{12})] 
  - \frac{v_3^2}{2}[\lambda_{23} + \lambda'_{23} + {\text{Re}}(\lambda^{\prime\prime}_{23})], \\
 m^2_{33} &=& \frac{v_1}{v_3}{\text{Re}}(m^2_{13}) + \frac{v_2}{v_3}{\text{Re}}(m^2_{23}) - \frac{v_3^2}{2}\lambda_3 \nonumber \\
&&- \frac{v_1^2}{2}[\lambda_{13} + \lambda'_{13} + {\text{Re}}(\lambda^{\prime\prime}_{13})] 
  - \frac{v_2^2}{2}[\lambda_{23} + \lambda'_{23} + {\text{Re}}(\lambda^{\prime\prime}_{23})].
\eea
\end{subequations}

\subsection{Charged Higgs Sector}

Since 
 CP-violation in the charged Higgs sector appears from the mixing between the physical charged Higgs mass eigenstates and gauge eigenstates $\phi_i^+$ ($i=1,2,3$), we could define a mixing matrix $U$, which is (following the notation of Ref.~\cite{Cree2011}):
\begin{eqnarray}
	\left( \begin{array}{c}
	\phi_1^+ \\ \phi_2^+ \\ \phi_3^+ \end{array} \right) 
	= U^{\dagger} 
	\left( \begin{array}{c}
	G^+ \\ H_2^+ \\ H_3^+ \end{array} \right),
	\label{eq:Udefinition}
\end{eqnarray}
where $G^+$ is the charged Goldstone boson. Here, $H_2^+$ and $H_3^+$ are the physical charged Higgs mass eigenstates.   The mixing matrix $U$ can be obtained through charged Higgs mass-squared matrix diagonalization, $V \supset \phi_i^- (\mathcal{M}^2_{H^\pm})_{ij} \phi_j^+$, where~\cite{Cree2011}:

\be
 \mathcal{M}^2_{H^\pm} = \left(
\begin{array}{ccc}
 \frac{v_2}{v_1}A_{12} + \frac{v_3}{v_1}A_{13} & -A_{12}+i B & -A_{13}-i\frac{v_2}{v_3} B \\
 -A_{12}-i B & \frac{v_1}{v_2}A_{12} + \frac{v_3}{v_2}A_{23} & -A_{23}+i\frac{v_1}{v_3} B \\
 -A_{13}+i\frac{v_2}{v_3} B & -A_{23}-i\frac{v_1}{v_3} B & \frac{v_1}{v_3}A_{13} + \frac{v_2}{v_3}A_{23} \\
\end{array}
\right),
\ee
with %
\begin{subequations}
\bea
 A_{12} &=& {\text{Re}}(m^2_{12}) - \frac{v_1v_2}{2}[\lambda'_{12} + {\text{Re}}(\lambda^{\prime\prime} _{12})],  \\ 
 A_{23} &=& {\text{Re}}(m^2_{23}) - \frac{v_2v_3}{2}[\lambda'_{23} + {\text{Re}}(\lambda^{\prime\prime}_{23})], \\ 
 A_{13} &=& {\text{Re}}(m^2_{13}) - \frac{v_1v_3}{2}[\lambda'_{13} + {\text{Re}}(\lambda^{\prime\prime}_{13})], \\ 
 B &=& -{\text{Im}}(m^2_{12}) + \frac{v_1v_2}{2}{\text{Im}}(\lambda ^{\prime\prime}_{12}). \label{eq:B_term}
\eea
Herein, the $B$ term will contain  CP-violation. Since we turn off CP-violation in the neutral Higgs sector through the condition in  Eq.~(\ref{EqAr:CPsplit}), Eq.~(\ref{eq:B_term}) becomes:
\bea
B = -\frac{v_1v_2}{2}\text{Im}(\lambda^{\prime\prime}_{12}).
\eea
\end{subequations} 
We then diagonalize the mass matrix of the physical charged Higgs bosons.  We perform it in two stages, starting with rotating to the Higgs basis using the rotation matrix
\begin{eqnarray}
	U_1 &=& \left( \begin{array}{ccc}
		\sin\gamma & 0 & \cos\gamma \\
		0 & 1 & 0 \\
		-\cos\gamma & 0 & \sin\gamma \end{array} \right)
		\left( \begin{array}{ccc}
		\cos\beta & \sin\beta & 0 \\
		-\sin\beta & \cos\beta & 0 \\
		0 & 0 & 1 \end{array} \right).
\end{eqnarray}
Here, the angles $\beta$ and $\gamma$ are given in terms of the VEVs and are defined as below:
\be\label{tanbetatangamma}
   \tan \beta = \frac{v_2}{v_1}, \qquad \tan \gamma = \frac{\sqrt{v^2_1 + v^2_2}}{v_3}.
 \ee
The charged Goldstone boson has been isolated by a rotation through the above $U_1$ matrix the charged Higgs boson mass matrix takes the following form:
\be
 \mathcal{M}^{\prime 2}_{H^\pm}
 = U_1\mathcal{M}^2_{H^\pm}U^\dagger_1
 =  \left(
  \begin{array}{ccc}
   0 & 0 & 0 \\
   0 & \mathcal{M}^{\prime 2}_{22} & \mathcal{M}^{\prime 2}_{23} \\
   0 & \mathcal{M}^{\prime 2*}_{23} & \mathcal{M}^{\prime 2}_{33} \\
  \end{array}
 \right), \label{Chmatrix}
\ee
where \cite{Cree2011}\footnote{ In Eq.~(A8) of Ref.~\cite{Cree2011}, there are two typos in the expression for $\mathcal{M}^2_{22}$ and we corrected in this work.}:
\begin{subequations}
\bea
 \mathcal{M}^{\prime 2}_{22} &=& \frac{v^2_{12}}{v_1v_2}A_{12} + \frac{v^2_2v_3}{v_1v^2_{12}}A_{13} + \frac{v_1^2v_3}{v_2v^2_{12}}A_{23}, \\
  \mathcal{M}^{\prime 2}_{33} &=& \frac{v_1v^2}{v_3v^2_{12}}A_{13} + \frac{v_2v^2}{v_3v_{12}^2}A_{23}, \\
  \mathcal{M}^{\prime 2}_{23} &=& \frac{v_2v}{v_{12}^2}A_{13} - \frac{v_1v}{v_{12}^2}A_{23} + i\frac{v}{v_3}B, 
\eea
\end{subequations}
and $v_{12}^2=v_1^2+v_2^2$. Then, we diagonalize the mass matrix in  Eq.~(\ref{Chmatrix}). We use the matrix $U_2$ to obtain the diagonalized result:
\be
 U_2 = \left(
  \begin{array}{ccc}
   1 & 0 & 0 \\
   0 & e^{-i\delta} & 0 \\
   0 & 0 & 1 \\
  \end{array} \right)
  \left(
  \begin{array}{ccc}
   1 & 0 & 0 \\
   0 & \cos\theta & \sin\theta e^{i\delta} \\
   0 & -\sin\theta e^{-i\delta} & \cos\theta \\
  \end{array} \right),
\ee
where the CP-violating phase parameter $\delta$ is defined as 
\be
 \delta = \text{phase}(\mathcal{M}^{\prime 2}_{23}),
\ee
with limit $0 \leq \delta < 2\pi$.  We pick the mixing angle $\theta$ in the range $-\pi/2 \leq \theta \leq 0$ for convenience\footnote{In the Democratic (or Type-Z) 3HDM, the coupling between $H_2^\pm$ and leptons becomes zero when $\theta = 0$. Similarly, there will be a zero value of the coupling between $H_3^\pm$ and leptons when $\theta = -\pi/2$.}, so that either $H_2^{\pm}$ or $H_3^{\pm}$ can be the lightest physical charged Higgs boson.  

The combination of the $U_1$ and $U_2$ matrix will give an explicitly full rotation matrix as the Eq.~(\ref{eq:Udefinition}) in  \cite{Cree2011}:
\begin{equation}
	U^{\dagger}
	= (U_2 U_1)^{\dagger}
	= \left( \begin{array}{ccc}
	s_\gamma c_\beta & -c_\theta s_\beta e^{i\delta} - s_\theta c_\gamma c_\beta  	&  s_\theta s_\beta e^{i\delta} - c_\theta c_\gamma c_\beta \\
	s_\gamma s_\beta 
		& c_\theta c_\beta e^{i\delta} - s_\theta c_\gamma s_\beta & -s_\theta c_\beta e^{i\delta} - c_\theta c_\gamma s_\beta  \\
c_\gamma	
		& s_\theta s_\gamma & c_\theta s_\gamma 
	\end{array} \right),
	\label{eq:Uexplicit}
\end{equation}
where $s = \sin$, $c = \cos$ for all mixing angle parameters ($\beta,\gamma,\theta,\delta$).  We will later use the $U^{\dagger}$ form for the Yukawa couplings rather than $U$, so that we will give the explicit parameter formulae in terms of $U^{\dagger}$.

\subsection{The Yukawa Lagrangian of the Charged Higgs Sector}

The Yukawa Lagrangian is:
\be\label{YukLag}
	\mathcal{L}_\text{Yukawa} = - \{ \bar{Q}_L\Phi_1\mathcal{G}_dd_R + \bar{Q}_L\tilde{\Phi}_2\mathcal{G}_uu_R + \bar{L}_L\Phi_3\mathcal{G}_l l_R + \text{h.c} \},
\ee
where $\tilde \Phi$ is the conjugate doublet form of $\Phi$ and is defined by $i \sigma^2 \Phi^*$. The fermion mass matrices $\mathcal{M}_f$ determine the Yukawa matrices $\mathcal{G}_f$ in the following form: $\mathcal{M}_f = \mathcal{G}_f v_i /\sqrt{2}$.

The Yukawa Lagrangian form of the charged Higgs boson sector is given by~\cite{Grossman:1994jb}:
\bea \label{eq:charge_lag}
 \mathcal{L}^\text{charged}_\text{Yukawa} &=& -\frac{\sqrt{2}}{v} \left\{ [X_2\bar{u}_LV \mathcal{M}_dd_R + Y_2\bar{u}_R\mathcal{M}_uVd_L + Z_2\bar{\nu}_L \mathcal{M}_ll_R]H^+_2 \right. \\ \nonumber
 	&+& \left. [X_3\bar{u}_LV \mathcal{M}_dd_R + Y_3\bar{u}_R\mathcal{M}_uVd_L + Z_3\bar{\nu}_L\mathcal{M}_ll_R]H^+_3 + \text{h.c.} \right\},
\eea
where $V$ is the SM CKM matrix. 
Since we adopt the NFC hypothesis, each Yukawa matrix associated to a charged Higgs boson is proportional to 
the mass matrix of the relevant fermion.
Thus the Yukawa terms include only six new parameters $X_i$, $Y_i$ and $Z_i$ $(i=2,3)$, which are determined by the charged 
Higgs mixing matrix $U^{\dagger}$ in Eq.~(\ref{eq:Uexplicit}) as
\be
	X_i = \frac{U^\dagger_{1i}}{U^\dagger_{11}}, \qquad Y_i = -\frac{U^\dagger_{2i}}{U^\dagger_{21}}, \qquad Z_i = \frac{U^\dagger_{3i}}{U^\dagger_{31}}.
	\label{XYZ}
\ee
These expressions are  for the Democratic 3HDM only while other Types will have a slightly different form.  The Yukawa couplings $X_i,Y_i,Z_i$ for other 3HDM types are listed in Tab.~\ref{tab:couplingfactors}.

\begin{specialtable}[H]
	\caption{The terms $X_i$, $Y_i$ and $Z_i$ are the Yukawa coefficients in  Eq.~(\ref{eq:charge_lag}) and are responsible for the coupling between $H_i^+$ to down-type quarks, up-type quarks and charged leptons, respectively. Here, $i$ equals to 2 and 3 as per the corresponding charged Higgs boson labels. The matrix $U^{\dagger}$ in each fraction is defined from Eq.~(\ref{eq:Uexplicit}).}
	\label{tab:couplingfactors}
	\centering
	\begin{tabular}{c ccc}
		\hline \hline
		Model & $X_i$ & $Y_i$ & $Z_i$ \\
		\hline
Type-I & $U^\dagger_{2i}/U^\dagger_{21}$ & $-U^\dagger_{2i}/U^\dagger_{21}$ & $U^\dagger_{2i}/U^\dagger_{21}$ \\
Type-II & $U^\dagger_{1i}/U^\dagger_{11}$ & $-U^\dagger_{2i}/U^\dagger_{21}$ & $U^\dagger_{1i}/U^\dagger_{11}$ \\
Type-X (Lepton-specific) & $U^\dagger_{2i}/U^\dagger_{21}$ & $-U^\dagger_{2i}/U^\dagger_{21}$ & $U^\dagger_{1i}/U^\dagger_{11}$ \\
Type-Y(Flipped) & $U^\dagger_{1i}/U^\dagger_{11}$ & $-U^\dagger_{2i}/U^\dagger_{21}$ & $U^\dagger_{2i}/U^\dagger_{21}$ \\
Type-Z (Democratic) & $U^\dagger_{1i}/U^\dagger_{11}$ & $-U^\dagger_{2i}/U^\dagger_{21}$ & $U^\dagger_{3i}/U^\dagger_{31}$ \\
	\hline \hline
	\end{tabular}
\end{specialtable}

{In our research, we assumed that all the extra neutral scalars (e.g., $H_{2,3}$ and $A_{2,3}$) are heavier than the masses of charged Higgs bosons. We have also taken the alignment limit so that the lightest CP-even neutral Higgs boson with mass 125 GeV has tree-level couplings identical to the SM one\footnote{For a review and discussions, see Ref.~\cite{Gunion:2002zf}}. In other words, as intimated, we limit the input parameters to six in total (and thus focus only on  charged Higgs sector physics):}

\[ M_{H^\pm_2}, M_{H^\pm_3}, \tan\beta, \tan\gamma, \theta, \delta. \]
In the above list, aside from the mass terms of the two physical charged Higgs boson states $M_{H^\pm_2}, M_{H^\pm_3}$, two of the other four mixing parameters, $\tan\beta$ and $\tan\gamma$, are correlated to the VEVs while $ \theta$ and $ \delta$ are the new free parameters in the Yukawa couplings. The first parameter is the mixing angle between two physical states whereas  the second is the CP-violation phase.  Our research focuses on the Democratic 3HDM, specifically, to investigate the relationship between these Yukawa coupling parameters and charged Higgs masses in CP-violating observables.

\subsection{Direct CP-asymmetry in $\bar{B} \to X_s \gamma$ and $\bar{B} \to X_{s+d} \gamma$}

The experimental measurement of the inclusive decay $\bar{B} \to X_s \gamma$ has been performed with two different methods. One is the fully-inclusive method and the other is the sum-of-exclusives method (also known as ``semi-inclusive''). The fully inclusive method consists on selecting one photon signal with a $B$ or $\bar{B}$ meson in a $B\bar{B}$ event via the decay $b \to s/d \gamma$\footnote{The symbol $B$ means $B^+$ or $B^0$ (containing $b$-antiquarks) while $\bar{B}$ denotes $B^-$ or $B^0$ (containing $b$-quarks). The symbol $X_s$ means any hadronic final state that originates from a $s$-quark hadronising (e.g., states with at least one kaon meson), $X_d$ means any hadronic final state that originates from a $d$-quark hadronising (e.g., states with at least one pion meson) while $X_{s+d}$ denotes any hadronic final state that is either $X_s$ or $X_d$.}. 
Simultaneously,  from the other $\bar{B}$ or $B$ meson (the ``tagged $B$ meson''), one singles out either a lepton ($e$ or $\mu$) or else a full reconstruction (hadronic or leptonic, called leptonic tagging or hadronic tagging, respectively) can be carried out. 
The fully inclusive BR($\bar{B} \to X_{s+d} \gamma$) has been measured by three collaboration groups. They are the CLEO  \cite{CLEO:2001gsa},  BaBar  \cite{BaBar:2012idb,BaBar:2012fqh} and  BELLE  \cite{Belle:2016ufb} collaborations. 

In the sum-of-exclusives approach, the selection criteria rules are sensitive to as many exclusive decays as possible in the hadronic final states $X_s$ and $X_d$ of the signal $\bar{B}$, plus it also requires a photon from the $b \to s/d + \gamma$ decay. 
In a $B\bar{B}$ event, in contrast, there is no other $B$ meson to tag.
In addition, the sum-of-exclusives method is sensitive to  $b \to s/d + \gamma$ de cays and results in the measurement of $\bar{B} \to X_s \gamma$ or $\bar{B} \to X_d \gamma$ observables separately whereas in the former method the contribution of $\bar{B} \to X_d \gamma$ events has to be substracted to obtain  measurements of $\bar{B} \to X_s \gamma$ observables alone. 
From both approaches, the experimental BR measurement of $\bar{B} \to X_s \gamma$ decays with a photon energy cut off between 1.7 and 2 GeV has been given, the most recent value of which and the corresponding theoretical prediction  can be seen
together  in Eqs.~(\ref{eq:bsgexp}) and~(\ref{eq:SMbsg}) of Appendix~\ref{appendix:a}. Our numerical analysis of $\bar{B} \to X_s \gamma$ is extrapolated from the Aligned 2HDM (A2HDM)~\cite{Borzumati1998}, with the effective Wilson coefficients running from the EW scale $\mu_W$ to the $b$-quark scale $\mu_b$. The formulae are given in Eqs. (\ref{eq:c07}) to (\ref{eq:bsg_final}) in Appendix~\ref{appendix:a}.

According to Eq.~(\ref{eq:bsgexp}) and~(\ref{eq:SMbsg}), it is clear that the BR($\bar{B} \to X_s \gamma$) measurements alone will not provide any evidence of new physics with BELLE II data, in contrast, the direct CP-asymmetry of this decay would possibly illustrate some new physics hints. 
The direct CP-asymmetry observables for $\bar{B} \to X_s \gamma$ and $\bar{B}\to X_{d} \gamma$ are defined as follows~\cite{Kagan:1998bh}:
\begin{equation}
\mathcal{A}_{X_{s(d)} \gamma}=\frac{\Gamma(\overline B\to X_{s(d)}\gamma)-\Gamma( B\to X_{s(d)}\gamma)}{\Gamma(\overline B\to X_{s(d)}\gamma)+\Gamma( B\to X_{s(d)}\gamma)}\,.
\end{equation} 
Here, $\mathcal{A}_{X_{s(d)} \gamma}$ (the short-distance contribution only) was firstly calculated through the Wilson coefficients in Ref.~\cite{Kagan:1998bh} and it was later combined with the long-distance contributions (also called ``resolved photons'') in Ref~\cite{Benzke:2010tq}. In our analysis, we took the following approximate form:
\bea \label{acpshortlong}
\mathcal{A}_{X_{s(d)}\gamma} &\approx & \pi \bigg \{  \bigg [ \bigg ( \frac{40}{81} - \frac{40}{9} \frac{\Lambda_c}{m_b} \bigg ) \frac{\alpha_s}{\pi} + \frac{\tilde{\Lambda}^{c}_{17}}{m_b} \bigg ] \text{Im} \frac{C_1}{C_{7\gamma}}  \\ \nonumber &-& \bigg ( \frac{4\alpha_s}{9\pi} - 4\pi \alpha_s e_{\text{spec}} \frac{\tilde{\Lambda}_{78}}{m_b}  \bigg ) \text{Im} \frac{C_{8g}}{C_{7\gamma}}\\  \nonumber & - & \bigg ( \frac{\tilde{\Lambda}^{u}_{17} - \tilde{\Lambda}^{c}_{17} }{m_b} + \frac{40}{9} \frac{\Lambda_c}{m_b} \frac{\alpha_s}{\pi} \bigg) \text{Im} \bigg( \epsilon_{s(d)} \frac{C_1}{C_{7\gamma}} \bigg ) \bigg \}\,.
\eea
Eq. (\ref{acpshortlong}) expresses the four asymmetries ($\mathcal{A}^{\pm}_{X_{s,d} \gamma}$ and $\mathcal{A}^{0}_{X_{s,d} \gamma} $ ) which are obtained from the values of $e_{\text{spec}}$ (the charge of the spectator quark) and $\epsilon_{s(d)}$ defined in Tab.~\ref{asymmetries}.
\begin{specialtable}[H] 
	\begin{center}
\caption{The choices of $e_{\text{spec}}$ and $\epsilon_{s(d)}$ in the generic formula for $\mathcal{A}_{X_s \gamma}$ that give rise to the four asymmetries. $e_{\text{spec}}$ is the spectator quark electric charge and neutral $B$ would be $-\frac{1}{3}$ and charged $B$ would give value of $\frac{2}{3}$ in the table.\label{asymmetries}}
\begin{tabular}{|c||c|c|}
\hline
$\mathcal{A}_{X_{s(d)} \gamma}$ &  $e_{\text{spec}}$ & $\epsilon_{s(d)}$  \\ \hline
$\mathcal{A}^0_{X_s \gamma}$ &  $-\frac{1}{3}$ & $\epsilon_{s}$  \\ 
$\mathcal{A}^\pm_{X_s \gamma}$ & $\frac{2}{3}$ &$\epsilon_{s}$  \\
$\mathcal{A}^0_{X_d \gamma}$ & $-\frac{1}{3}$ & $\epsilon_{d}$   \\
$\mathcal{A}^\pm_{X_d \gamma}$ & $\frac{2}{3}$ & $\epsilon_{d}$  \\
\hline
\end{tabular}
	\end{center}
\end{specialtable}
The magnitudes of the
long-distance or non-perturbative contribution are determined by the $\tilde{\Lambda}^{u}_{17} , \tilde{\Lambda}^{c}_{17} $ and $ \tilde{\Lambda}_{78}$ hadronic parameters. Their updated allowed ranges in \cite{Gunawardana:2019gep} are as follows:
\bea 
-660  \; \text{MeV} < \tilde{\Lambda}^{u}_{17} < + 660  \;\text{MeV}\,,\\ \nonumber
-7  \; \text{MeV} < \tilde{\Lambda}^{c}_{17} <  + 10  \; \text{MeV} \,,\\ \nonumber
17  \;\text{MeV} <  \tilde{\Lambda}_{78} <190  \; \text{MeV}\,.
\label{hardonic_parameter}
\eea
The short-distance contributions to $\mathcal{A}_{X_{s(d)}\gamma}$ are the terms that are independent of hadronic parameters and if long-distance
terms are neglected, $\mathcal{A}^0_{X_{s(d)}\gamma}=\mathcal{A}^\pm_{X_{s(d)}\gamma}$.
Other parameters are used as follows: $\Lambda_c = 0.38 \; \text{GeV}$,
$\epsilon_s = (V_{ub} V^*_{us})/ (V_{tb}V^*_{ts}) = \lambda^2 (i\bar{\eta} - \bar{\rho})/ [1- \lambda^2 (1 - \bar{\rho} + i \bar{\eta})]$ (in terms of Wolfenstein parameters) and $\epsilon_d = (V_{ub} V^*_{ud})/ (V_{tb}V^*_{td}) = (\bar{\rho}-i\bar{\eta})/ (1- \bar{\rho} + i \bar{\eta})$. The $C_i$ terms are relative Wilson coefficients. (Notice that $C_1$ is referred to as $C_2$ in \cite{Kagan:1998bh}.)
The Wilson coefficients are real in the SM and the only non-zero term in $\mathcal{A}_{X_{s(d)}\gamma}$ is the term with $\epsilon_{s(d)}$. For the imaginary parts, one has ${\rm Im}(\epsilon_d)/{\rm Im}(\epsilon_s)\approx -22$,  $\epsilon_s$ being of order 
 $\lambda^2$ while $\epsilon_d$ is of order 1.
For the short-distance contribution only (i.e., when $(\Lambda^{u}_{17}-\Lambda^{c}_{17})/{m_b}$ terms are neglected in Eq.~(\ref{acpshortlong})),  one has
$\mathcal{A}_{X_{s}\gamma}\approx 0.5\%$ and $\mathcal{A}_{X_{d}\gamma}\approx 10\%$. Thus, the prediction of this observable suggests that BSM would possibly be probed through the imaginary parts of the Wilson coefficients. However, the prediction of the SM value for $\mathcal{A}_{X_{s}\gamma}$ is increased to $-1.9\% < \mathcal{A}_{X_{s}\gamma} < 3.3\%$ in 
\cite{Gunawardana:2019gep} from $-0.6\% < \mathcal{A}_{X_{s}\gamma} < 2.8\%$ from Ref.~\cite{Benzke:2010tq} due to the update of the long distance contribution, so this does not seem to be a very powerful new physics probe and thus  another observable, the difference of CP-asymmetries for the charged and neutral $B$ mesons,
$\Delta\mathcal{A}_{X_{s}\gamma}=\mathcal{A}^\pm_{X_{s}\gamma}-\mathcal{A}^0_{X_{s}\gamma}$ was proposed, which is given by~\cite{Benzke:2010tq}:
\bea \label{eq:delta_cp}
\Delta\mathcal{A}_{X_{s}\gamma} \approx 4 \pi^2 \alpha_s \frac{\tilde{\Lambda}_{78}}{m_b} \text{Im} \frac{C_{8g}}{C_{7\gamma}}\,.
\eea
The above formula can be obtained from Eq.~(\ref{acpshortlong}) on  the condition that there is no cancellation with terms containing $e_{\text{spec}}$.  
In the SM, the prediction of this observable is essentially zero due to real Wilson coefficients  only. Thus, it is potentially a more effective probe than the observable $\mathcal{A}_{X_{s(d)}\gamma}$. Note that $\Delta\mathcal{A}_{X_{s}\gamma}$ depends on three parts: the long-distance hadronic parameter $\tilde{\Lambda}_{78}$ and two short-distance Wilson coefficients $C_{7\gamma},C_{8g}$. Thus, the uncertainty of the long-distance term would somehow affect the measured value. 

An even more interesting observable, called the untagged (fully inclusive)-asymmetry, is given by:
\bea
\mathcal{A}_{\rm CP} (\bar{B} \to X_{s+d} \gamma) = \frac{(\mathcal{A}^0_{X_s \gamma} + r_{0\pm}\mathcal{A}^{\pm}_{X_s \gamma})  + R_{ds} (\mathcal{A}^0_{X_d \gamma} + r_{0\pm}\mathcal{A}^{\pm}_{X_d \gamma})}{(1 + r_{0\pm}) (1 + R_{ds})}\,,
\eea
where $R_{ds}$ is the ratio {\rm BR}$(B\to d\gamma)$/{\rm BR}$(B\to s\gamma)\approx |V_{td}/V_{ts}|^2$ and the parameter $r_{0\pm}$ is
\bea
r_{0\pm} = \frac{N^+_{X_s} + N^-_{X_s}}{N^{\bar{0}}_{X_s} + N^0_{X_s}},
\eea
with $N^+_{X_s}$ being the number of $B^+$ mesons that decay to $X_s \gamma$ and similar decays (e.g., $B^-,\bar{B^0}, B^0$). We take the value $r_{0\pm} = 1$ in our numerical analysis (the experimental value is approximately 1.03 \cite{Belle-II:2018jsg}). The SM prediction of this observable is zero even with long-distance/non-perturbative terms \cite{Soares:1991te} and confirms $\mathcal{A}_{\rm CP} (\bar{B} \to X_{s+d} \gamma)$ to be a cleaner test for new physics. In this work, we study these three observables, $\mathcal{A}_{X_{s(d)}\gamma}$, $\Delta\mathcal{A}_{X_{s}\gamma}$ and $\mathcal{A}_{\rm CP} (\overline B \to X_{s+d} \gamma), $ based on the charged Higgs contributions to the relative Wilson coefficients in the 3HDM.

\subsection{EDMs from Charged Higgs Bosons in the 3HDM}

\begin{figure}[H]
\centering
\includegraphics[scale=0.4]{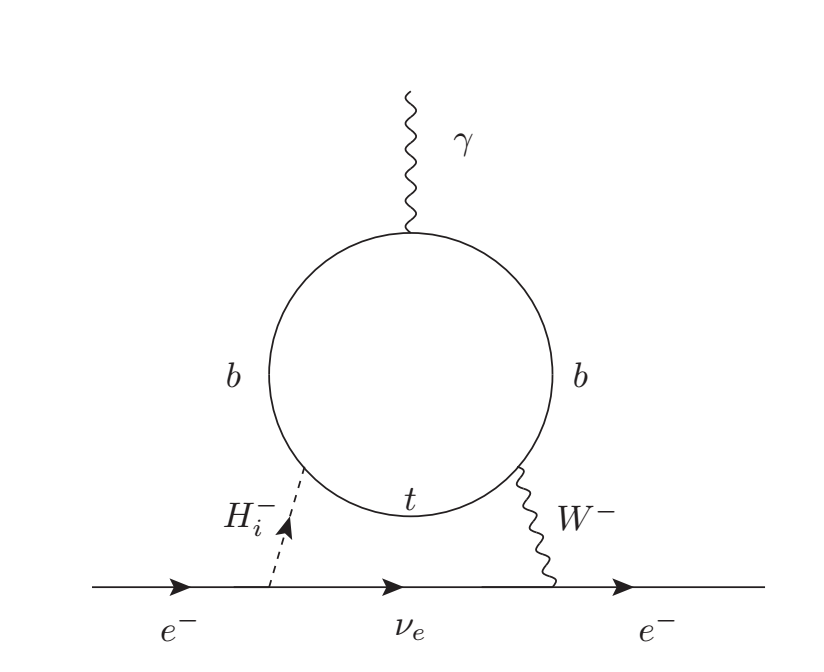}
\caption{In the 3HDM, the typical Barr-Zee type diagram contributing to the eEDM and contains the  charged  ($H^\pm_i$, $i = 2,3$) Higgs bosons. Reproduced from Ref.~\cite{Logan:2020mdz}.}
\label{fig:barrzeetb}
\end{figure}

\begin{figure}[H]
\centering
\includegraphics[scale=0.5]{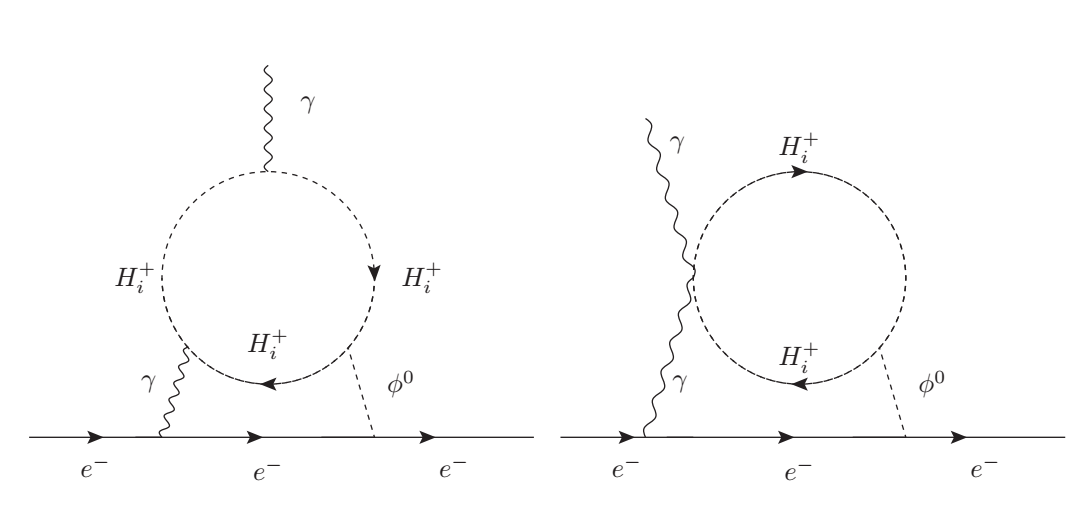}
\caption{In the 3HDM, two types of Barr-Zee diagrams that contribute to the eEDM and contain neutral ($\phi^0$) and charged ($H_i^\pm$, $i=2,3$) Higgs bosons. Reproduced from Ref.~\cite{Logan:2020mdz}.}
\label{fig:barrzeeHp}
\end{figure}

This section discusses the contributions of the charged Higgs bosons of the 3HDM to both eEDM and nEDM. 
For this purpose, we take the results available in the literature for the 2HDM and make a straightforward generalisation of these.
In the case of the eEDM, the contribution comes from the CP-violating phases in the Yukawa couplings of charged Higgs bosons to fermion pairs. The tiny electron Yukawa coupling suppresses the charged Higgs one-loop contribution and becomes subdominant. The two-loop Barr-Zee type contribution is dominant (the diagram is shown in Fig.~\ref{fig:barrzeetb}) and it was calculated in Ref.~\cite{BowserChao1997} within the 2HDM for the first time (see also Ref.~\cite{Jung2014}). The charged Higgs components also appear in the  Barr-Zee type diagrams of Fig.~\ref{fig:barrzeeHp}, where $\phi^0$ is any of the neutral scalars in the model. 
It has been noted that, in the context of the A2HDM~\cite{Kanemura2020}\footnote{A 2HDM which prevents tree-level FCNC by special Yukawa alignment in flavour space rather than general 2HDM which imposed a condition that only one scalar field couples to one fermion type field\cite{Pich:2009sp}.},  the diagrams of Fig.~\ref{fig:barrzeeHp} can have a significant contribution and lead to cancellations with the diagrams of Fig.~\ref{fig:barrzeetb}. However, as the 3HDM we study here has a neutral sector with no CP-violating phases, the diagrams of Fig.~\ref{fig:barrzeeHp} do not contribute to the eEDM because the $\phi^0 ee$ and $\phi^0 H_i^+ H_i^-$ couplings contain no CP-violating phase\footnote{The latter one could not have a CP-violating phase because the coefficient would be real due to the Hermitian form of  the Lagrangian where it appears.}. (Also, we have assumed our neutral Higgs sector to be heavy and decoupled, apart from the needed SM-like Higgs state.)
There are non-trivial phases in the couplings $\phi^0 H_2^+ H_3^-$, but they do not appear in the diagram of Fig.~\ref{fig:barrzeeHp} because the charged Higgs boson couples with a photon diagonally. 

With the choice of a CP-conserving case in the neutral Higgs sector (only the charged Higgs sector contains the CP-violating phase), the dominant Barr-Zee type contribution of the charged Higgs boson to the eEDM in the 2HDM~\cite{BowserChao1997,Jung2014} can be extrapolated to the 3HDM as follows:
\begin{eqnarray}
\frac{d_e(M_{H^{\pm}_2},M_{H^{\pm}_3})_{BZ}}{2} &=& - m_e \frac{12 G^2_F M^2_W}{(4\pi)^4} |V_{tb}|^2   \nonumber  \\
&\times & \bigg[{\text{Im}} (- Y_2^*Z_2) \left(q_t F_t(\textit{z}_{H^{\pm}_2},\textit{z}_W) + q_b F_b(\textit{z}_{H^{\pm}_2},\textit{z}_W)\right)   \nonumber \\
&+& {\text{Im}} (- Y_3^*Z_3) \left(q_t F_t(\textit{z}_{H^{\pm}_3},\textit{z}_W) + q_b F_b(\textit{z}_{H^{\pm}_3},\textit{z}_W)\right) \bigg],
\label{eq: deforH}
\end{eqnarray}
where $q_t,q_b = 2/3,-1/3$ are the top and bottom quark electric charges, respectively, while $z_{a}=M_a^2/m_t^2$.
\bea
F_q(\textit{z}_{H^{\pm}_i},\textit{z}_W) &=& \frac{T_q(\textit{z}_{H^{\pm}_i}) - T_q(\textit{z}_W) }{\textit{z}_{H^{\pm}_i} - \textit{z}_W},    \\
T_t(\textit{z}) &=& \frac{1- 3\textit{z}}{\textit{z}^2} \frac{\pi^2}{6} + \bigg(\frac{1}{\textit{z}} - \frac{5}{2}\bigg) \text{log}\textit{z} -\frac{1}{\textit{z}} - \bigg(2 - \frac{1}{\textit{z}}\bigg) \bigg(1 - \frac{1}{\textit{z}}\bigg) \text{Li}_2 (1 -\textit{z}),  \nonumber \\
T_b(\textit{z}) &=& \frac{2\textit{z} - 1}{\textit{z}^2} \frac{\pi^2}{6} + \bigg(\frac{3}{2} - \frac{1}{\textit{z}}\bigg) \text{log}\textit{z} + \frac{1}{\textit{z}} - \frac{1}{\textit{z}} \bigg(2 - \frac{1}{\textit{z}}\bigg) \text{Li}_2 (1 -\textit{z}). \nonumber
\eea
In Ref.~\cite{BowserChao1997}, the calculation was done with $m_b = 0$ so that the only contribution will come from the top-quark Yukawa couplings $m_t Y_i/v$. 
The contribution terms with $m_b X_i/v$ could be introduced via a non-zero bottom mass, which could affect the results and increase  the importance of large $\tan\beta$ values.  
At last, we believe other purely fermionic or gauge bosons eEDM contributions \cite{Pospelov1991,Chang1991} at the loop level to be negligible against the current experimental bound since they remain identical to those in the SM.

\begin{figure}[H]
\includegraphics[scale=0.4]{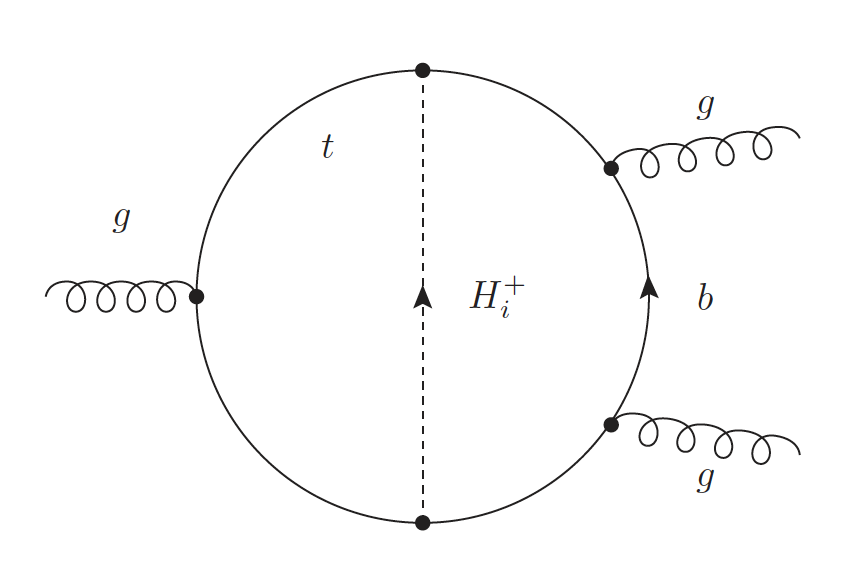}
\includegraphics[scale=0.4]{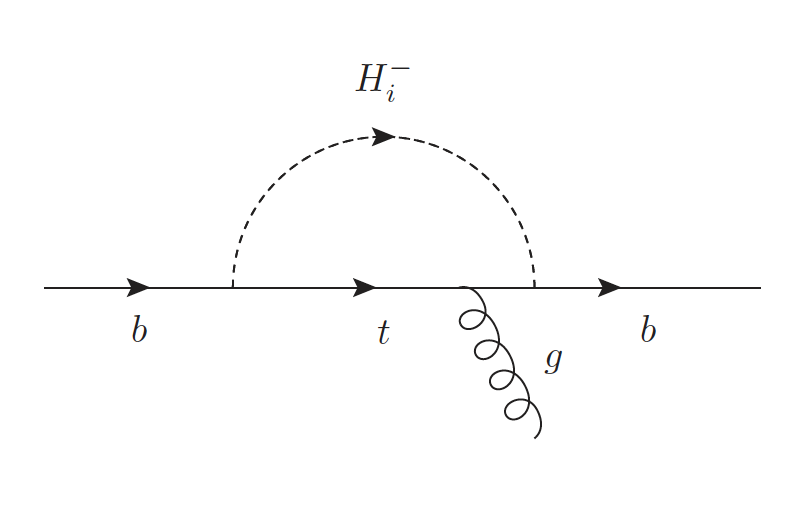}
\caption{{Left panel}:  The Weinberg operator contribution from two-loop charged Higgs boson diagrams. 
{Right panel}: The bottom quark CEDM contribution from  one-loop charged Higgs boson diagrams. Reproduced from Ref.~\cite{Logan:2020mdz}.  }
\label{fig:bCEDM}
\end{figure}

To measure the nEDM CP-violation from charged Higgs boson exchange, we need to take a variety of Effective Field Theory (EFT) operators. In the context of the 2HDM, Jung and Pich~\cite{Jung2014} identified three types of these: 1) the up- and down-type quarks contributing  to four-fermion operators are induced by CP-violating Higgs exchange; 2) the Weinberg operator (the CP-violating three-gluon operator) could be large since there is no suppression from either quark masses or CKM matrix elements; 3) the up- and down-type quark contribution to EDMs and Chromo-Electric Dipole Moments (CEDMs) involve Barr-Zee type two-loop diagrams. 
The contributions from the four fermion operators and the up- and down-type quark CEDMs are suppressed by the light quark masses, so that the dominant contribution will come from the Weinberg operator. The charged Higgs contribution to the Weinberg operator can be seen on the left of Fig.~\ref{fig:bCEDM}. We compute the contribution using the effective field theory (EFT) approach, the computation of which includes only the one-loop short-distance part at the high scale $\mu_{tH} = \mu_{m_t}$. We show this bottom-quark CEDM at the right panel of Fig.~\ref{fig:bCEDM}.

The contribution of the Weinberg operator to the nEDM is given by~\cite{Jung2014}:
\begin{equation}\label{formula1}
|d_n(C_W)/e|  = \left[1.0  {+1.0 \atop -0.5} \right] \times 20\,  \text{MeV} \,  C_W(\mu_h).
\end{equation}
In the above equation, the sign inside the square bracket is unknown, and there is a factor of 2 on the theoretical uncertainty magnitude.  In our numerical analysis, we use the central theoretical value from the discussion in Ref.~\cite{Jung2014}. At the hadronic scale $\mu_h \sim 1$~GeV, the Wilson coefficient $C_W$ is evaluated as:
\begin{equation}\label{formula2}
C_W(\mu_h) = \eta^{\kappa_W}_{c-h}  \eta^{\kappa_W}_{b-c} \bigg( \eta^{\kappa_W}_{t-b} C_W(\mu_{tH}) +  \eta^{\kappa_C}_{t-b} \frac{g^3_s(\mu_{b})}{8\pi^2 m_b} \frac{d^C_b(\mu_{tH})}{2} \bigg),
\end{equation}
where $C_W(\mu_{tH}) = 0$ and $\mu_{tH} \thicksim m_t$. The reason is that at the scale $m_t$  the short-distance contribution to the Weinberg operator does not involve the charged Higgs boson one. Here, 
$d_b^C(\mu_{tH})$ is the bottom quark CEDM short-distance contribution term and will be listed below.  These short-distance contribution terms run down to the lower scale $\mu_b = m_b$ from $\mu_{tH} = m_t$ by the factors of $\eta_{t-b} = \alpha_s(\mu_{tH})/\alpha_s(\mu_b)$ with the appropriate power $\kappa_i = \gamma_i/(2 \beta_0$), where $\gamma_W = N_C + 2 n_f$ and $\gamma_C = 10 C_F - 4 N_C$ are the leading order (LO) anomalous dimensions of the Weinberg  and  $b$-quark CEDM operator, respectively. The one-loop $\beta$ function of QCD is $\beta_0 = (11 N_C - 2 n_f)/3$.  In this formula, $N_C = 3$, $C_F = 4/3$ and $n_f$ is the active quark flavor number in the QCD running QCD at the appropriate scale (e.g.,  the scale between the top and bottom quark masses lead $n_f$ equals to 5 rather than 6.). At the lower scale $\mu_b$, the bottom quark is integrated out and other operators are matched, then the remaining Weinberg operator is further running down to the hadronic scale $\mu_h$ ($\approx 1$ GeV) by two steps (the charm quark is same as bottom quark and integrated at $\mu_c = m_c$), so that two more factors are rising, $\eta_{b-c}^{\kappa_W}$ and $\eta_{c-h}^{\kappa_W}$, are relevant. The running $\alpha_s(\mu)$ at LO is given by
\begin{equation}
\alpha_s(\mu) = \frac{\alpha_s(M_Z)}{v(\mu)},
\end{equation}
with 
\begin{equation}
v(\mu) = 1 - \beta_0 \frac{\alpha_s(M_Z)}{2\pi} \log \bigg(\frac{M_Z}{\mu} \bigg).
\end{equation}
Here, 
$M_Z$ is the $Z$ boson mass with the running coupling $\alpha_s(M_Z)$. Finally, the right panel of Fig.~\ref{fig:bCEDM} is the bottom quark CEDM with the high-scale one-loop charged Higgs boson contribution covered in the 2HDM~\cite{Jung2014}. The extrapolation to the 3HDM is then given by
\begin{eqnarray}
\frac{d^C_b(\mu_{tH})}{2} &=& - \frac{G_F}{\sqrt{2}} \frac{1}{16\pi^2}  |V_{tb}|^2 m_b(\mu_{tH}) 
	\left[{\text{Im}}(-X_2Y_2^*) x_{tH_2} \bigg(\frac{\log(x_{tH_2})}{(x_{tH_2} - 1)^3} + \frac{(x_{tH_2} - 3)}{2(x_{tH_2} - 1)^2}\bigg) \right. \nonumber \\
	&& \qquad \qquad \qquad \left. + {\text{Im}}(-X_3Y_3^*) x_{tH_3} \bigg(\frac{\log(x_{tH_3})}{(x_{tH_3} - 1)^3} + \frac{(x_{tH_3} - 3)}{2(x_{tH_3} - 1)^2}\bigg) \right],
\label{formula7}
\end{eqnarray}
where $x_{tH_i} = m_t^2 / M_{H^{\pm}_i}^2$.  Again, purely negligible fermions and gauge contributions \cite{Jung2014} give the same terms as in the SM.

\section{Results}\label{section3}

\subsection{CP-asymmetry Observables With Charged Higgs Boson Yukawa Couplings}

The measurement of three asymmetries, $\mathcal{A}^{\text{tot}}_{X_{s}\gamma}$, $\mathcal{A}_{\rm CP} (\overline B \to X_{s+d} \gamma) $ and $\Delta\mathcal{A}_{X_{s}\gamma}$ have all been carried out by the BELLE and BaBar collaboration, and their most recent measurements is presented in Tab. \ref{bsy_measurement}, as well as the world average which was given in \cite{ParticleDataGroup:2018ovx}. 
In the near future, the experiment BELLE-II will give more precise measurements (see Ref.~\cite{Belle-II:2018jsg} ). The present accumulated 74 fb$^{-1}$ integrated luminosity data is one-sixth of the BELLE experiment and one-tenth of the BaBar experiment. Tab.~\ref{bsy_prospects} has summarised the estimated precision with an integrated luminosity of 50 ab$^{-1}$ that is expected to be reached around the year 2030. Note that both leptonic- and hadronic-tag refer to the fully inclusive method discussed above.
In Tab.~\ref{bsy_prospects}, it is clear that the SM prediction of $\mathcal{A}_{\rm CP} (\overline B \to X_{s+d} \gamma) $ is zero, and thus a central value of 2.5\% with error 0.5\% would provide a 5$\sigma$ signal of physics beyond the SM. 

\begin{specialtable}[H]
\caption{Measurements (given as percentage values) of ${\mathcal{A}}^{\rm tot}_{X_s \gamma}$,  $\mathcal{A}_{\rm CP} (\overline  B \to X_{s+d} \gamma)$ and $\Delta\mathcal{A}_{X_{s}\gamma}$ at BELLE collaboration, BaBar collaboration and the
world average.\label{bsy_measurement}}
	\begin{small}
\begin{tabular}{c||ccc}
\hline
& BELLE &  BaBar & World average  \\ 
\hline
${\mathcal{A}}^{\rm tot}_{X_s \gamma}$ 
& $(1.44\pm 1.28\pm 0.11)\%$\cite{Belle:2018iff}
& $(1.73\pm 1.93\pm 1.02)\%$\cite{BaBar:2014czi} 
& $1.5\%\pm 1.1\%$\cite{ParticleDataGroup:2018ovx} \\
$\mathcal{A}_{\rm CP} (\overline B \to X_{s+d} \gamma)$ 
& $(2.2\pm 3.9 \pm 0.9)\%$\cite{Belle:2015vct} 
& $(5.7\pm 6.0\pm 1.8)\%$\cite{BaBar:2012idb} 
& $1.0\%\pm 3.1\%$\cite{ParticleDataGroup:2018ovx} \\
$\Delta\mathcal{A}_{X_{s}\gamma}$ 
& $(3.69\pm 2.65\pm 0.76)\%$\cite{Belle:2018iff}
& $(5.0\pm 3.9\pm 1.5)\%$\cite{BaBar:2014czi} 
& $4.1\%\pm 2.3\%$\cite{ParticleDataGroup:2018ovx} \\
\hline
\end{tabular}
	\end{small}
\end{specialtable}

\begin{specialtable}[H]
\caption{SM predictions of $\mathcal{A}^{\rm tot}_{X_s \gamma}$,  $\mathcal{A}_{\rm CP} (\overline B \to X_{s+d} \gamma)$ and $\Delta\mathcal{A}_{X_{s}\gamma}$ and each expected experimental error at BELLE II with integrated Luminosity at 50 ab$^{-1}$ \cite{Belle-II:2018jsg}. \label{bsy_prospects}}
	\begin{small}
\begin{tabular}{c||cccc}
\hline
& {\footnotesize SM Prediction} & {\footnotesize Leptonic tag}  & {\footnotesize Hadronic tag} & {\footnotesize Sum of exclusives} \\ 
\hline
$\mathcal{A}^{\rm tot}_{X_s \gamma}$ 
& $-1.9\% < \mathcal{A}_{X_s \gamma} < 3.3\%$
&  x  & x &  0.19\%\\
$\mathcal{A}_{\rm CP} (\overline B \to X_{s+d} \gamma)$ 
&  0 & 0.48\% & 0.70\% & 0.3\%   \\
$\Delta\mathcal{A}_{X_{s}\gamma}$ 
& 0 &  x & 1.3\% & 0.3\%   \\
\hline
\end{tabular}
	\end{small}
\end{specialtable}

In the following, we analyse the parameter space of the 3HDM and the charged Higgs contribution to each observable to contrast with the experimental measures. We take the $3\sigma$ experimental bound of each world average CP-asymmetry observable to be the plot limit and take the $3\sigma$ bound of $\bar{B} \to X_s \gamma$ (which is $2.87 \leq \bar{B} \to X_s \gamma \leq 3.77 \times 10^{-4}$) as the allowed region.

\begin{figure}[H]
\centering
\includegraphics[scale=0.46]{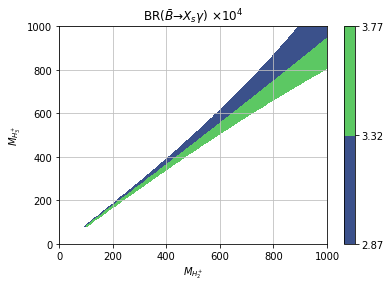}
\includegraphics[scale=0.46]{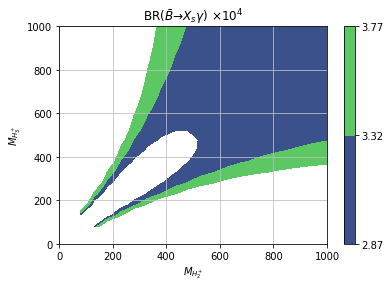}
\caption{BR($\bar{B} \to X_s \gamma$) in the plane [$M_{H^{\pm}_2},M_{H^{\pm}_3}$], with $\theta = - \pi/4, \tan\beta = 10, \tan\gamma = 1$. Left panel: $\delta = 0$. Right panel: $\delta = \pi/2$. Reproduced from Ref.~\cite{Akeroyd2020}.  \label{fig:bsg_m1m2_minpi4}}
\end{figure}

First, we show the effect of the CP-violating phase parameter in the BR($\bar{B} \to X_s \gamma$) over the plane of the two charged Higgs state masses. In Fig.~\ref{fig:bsg_m1m2_minpi4}, the magnitude of BR($\bar{B} \to X_s \gamma$) is saturated with mixing parameters $\theta = - \pi/4, \tan\beta = 10, \tan\gamma = 1$.  The left panel is the CP-conserving case, with a value of $\delta = 0$, while the right panel with is the maximal CP-violation scenario, with $\delta = \pi/2$. A non-zero value of the CP-phase $\delta$ significant increases the allowed parameter space in the plane [$M_{H^{\pm}_2},M_{H^{\pm}_3}$]. In this case, no CP-phase $\delta$ would lead more contribution to Yukawa couplings and affect the value of $\bar{B} \to X_s \gamma$ to be larger which could be seen from real combination components of $X_2Y^*_2,X_3Y^*_3$ in Appendix \ref{appendix:char}.

\begin{figure}[H]
\centering
\includegraphics[scale=0.46]{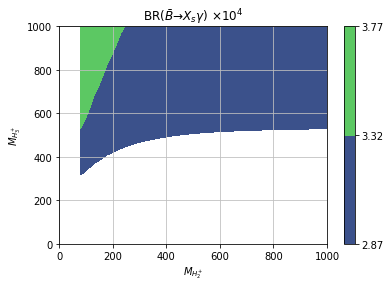}
\includegraphics[scale=0.46]{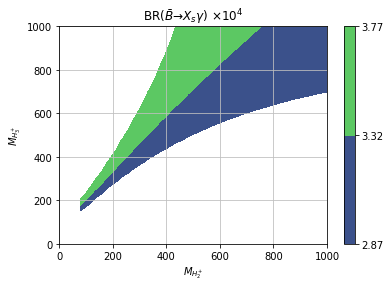}
\caption{BR($\bar{B} \to X_s \gamma$) in the plane [$M_{H^{\pm}_2},M_{H^{\pm}_3}$]. We fixed parameters $\theta = - \pi/2.1, \tan\beta = 10, \tan\gamma = 1$. Left panel: $\delta = 0$. Right panel: $\delta = \pi/2$. Reproduced from Ref.~\cite{Akeroyd2020}.  \label{fig:bsg_m1m2_minpi21}}
\end{figure}

In addition, Fig.~\ref{fig:bsg_m1m2_minpi21} shows the same parameter space as Fig.~\ref{fig:bsg_m1m2_minpi4} with a different value $\theta = - \pi/2.1$. Here, the non-zero phase has changed the surviving parameter space, and more solutions have occurred for charged Higgs lighter than 400 GeV, which can be seen from comparing the left panel to the right one.   Thus, it is clear that $\delta$ would play an essential role in the relative dynamics of the two charged Higgs bosons in the 3HDM.

\begin{figure}[H]
\centering
\includegraphics[scale=0.3]{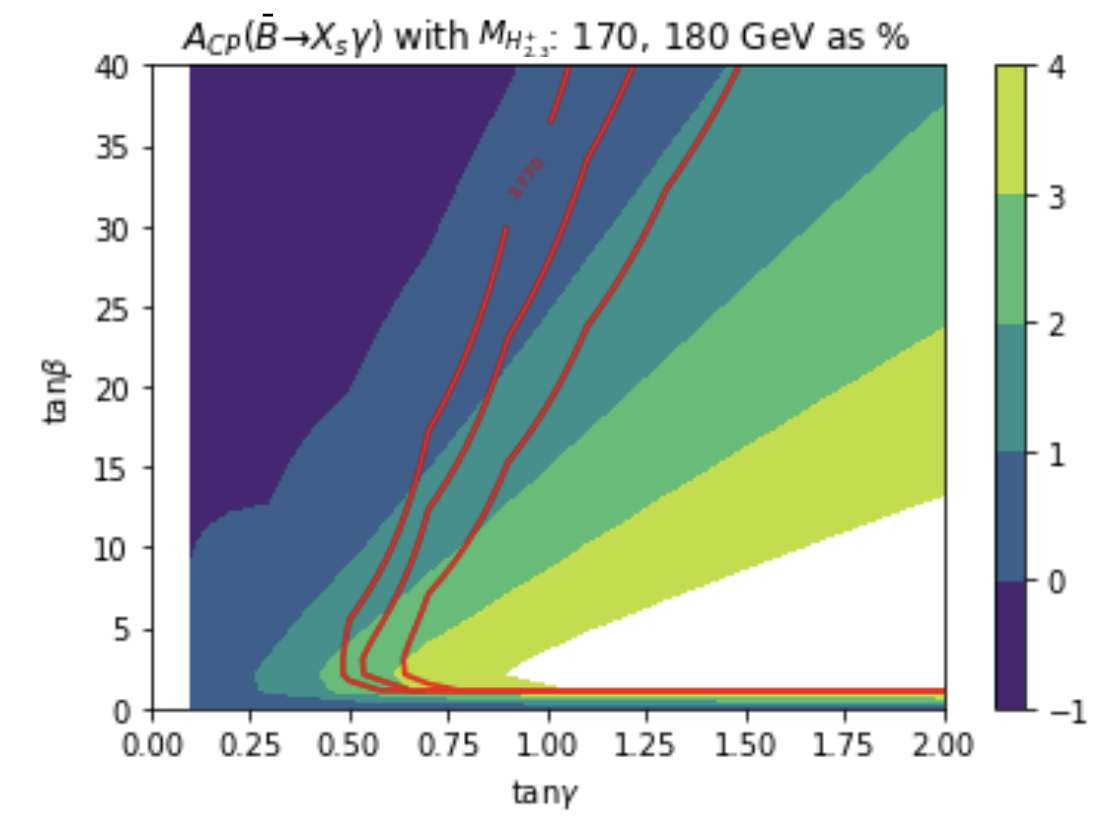}
\includegraphics[scale=0.3]{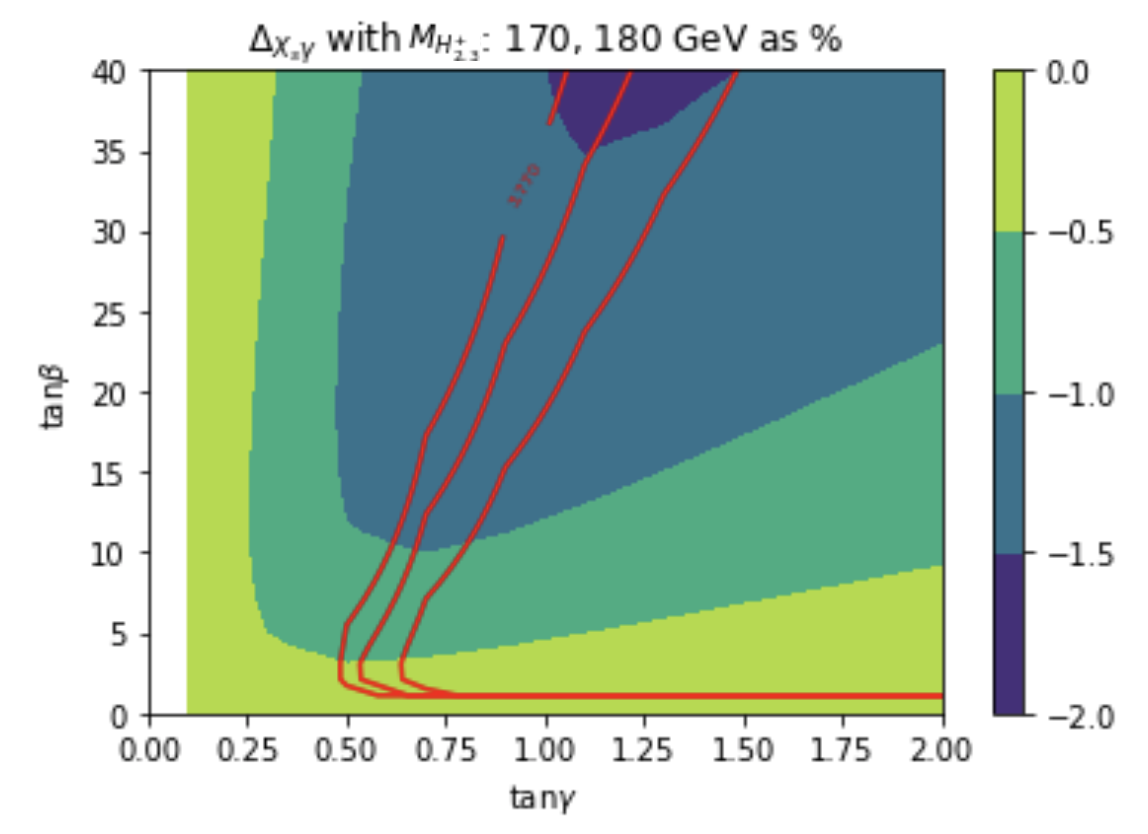}\\
\includegraphics[scale=0.3]{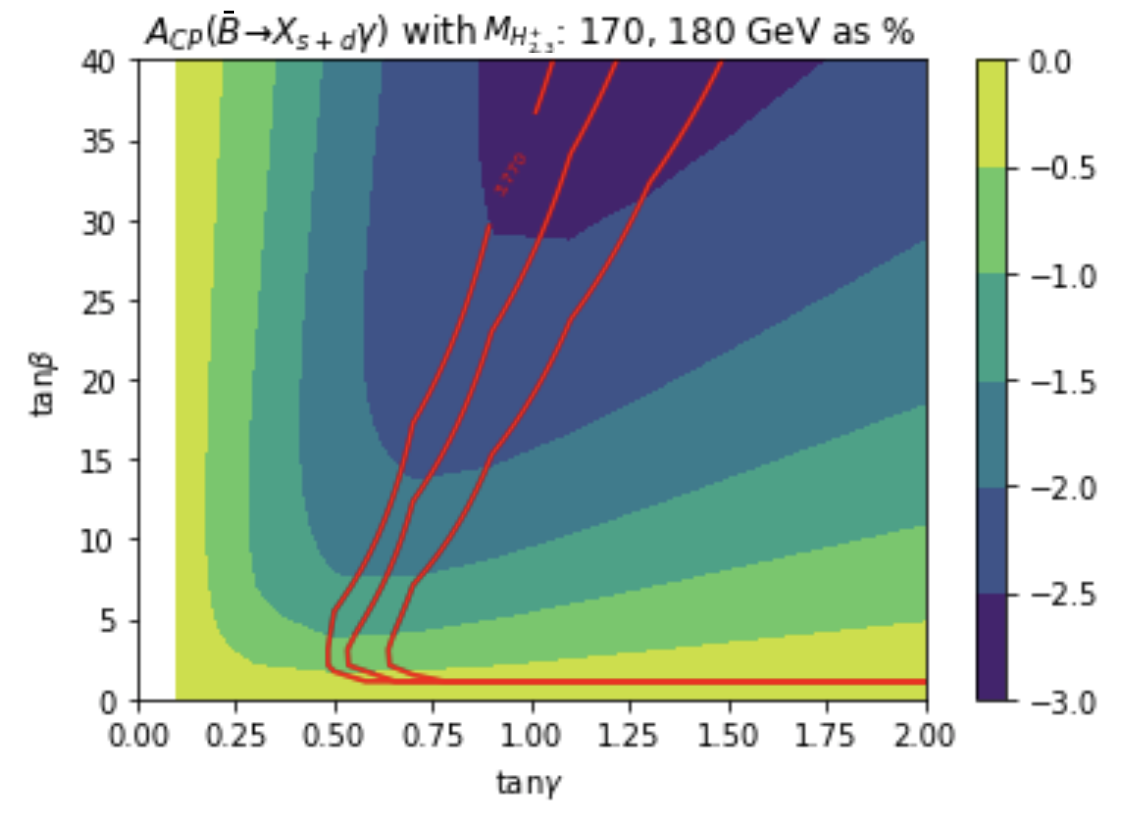}
\caption{CP-Asymmetries (as a percentage (\%)) in the plane [$\tan\gamma, \tan\beta$] with $M_{H^{\pm}_2} = 170$ GeV, $M_{H^{\pm}_3} = 180$ GeV, $\theta=-0.85$ and
    $\delta=2.64$.
    {Left top panel}: $\mathcal{A}_{\rm CP} (\bar{B} \to X_s \gamma )$. {Right top panel}:  $\Delta A_{\rm CP}$. {Bottom panel}: $\mathcal{A}_{\rm CP} (\overline B \to X_{s+d} \gamma)$. Reproduced from Ref.~\cite{Akeroyd2020}.  \label{bsg_170180_b_g} }  
\end{figure}

We now show one benchmark point which gives large values of CP-asymmetry results in the 3HDM. In Fig.~\ref{bsg_170180_b_g}, the observables $\mathcal{A}^{\text{tot}}_{X_{s}\gamma}$ (left top panel), $\Delta\mathcal{A}_{X_{s}\gamma}$ (Right top panel) and $\mathcal{A}_{\rm CP} (\overline B \to X_{s+d} \gamma) $ (bottom panel) are presented in the Yukawa coupling parameter plane [$\tan\gamma,\tan\beta$]. In these figures, the charged Higgs masses are fixed as $M_{H^{\pm}_2} = 170$ GeV, $M_{H^{\pm}_3} = 180$ GeV, $\theta=-0.85$ and $\delta=2.64$ and the long-distance (hadronic) parameters are taken to be $\tilde{\Lambda}^{u}_{17} = 0.66$ GeV, $\tilde{\Lambda}^{c}_{17} = 0.010$ GeV and $\tilde{\Lambda}_{78} = 0.19$ GeV. The scale $\mu_b$ is taken to be $m_b=4.8$ GeV. Three red lines of the figures are the allowed 3$\sigma$ BR($\bar{B} \to X_s \gamma$) bound, which is the experimental limit as Eq. (\ref{eq:bsgexp}) (from top to right) gives the upper (3$\sigma$) limit, central value, and the lower ($3\sigma$) limit. The white region of each plot is excluded (due to violation the current 3$\sigma$ world average experimental limit as in Table.~\ref{bsy_measurement}) by the certain CP-asymmetry observables (e.g., white region with $\tan\gamma > 1$ is ruled out in $\mathcal{A}^{\text{tot}}_{X_{s}\gamma}$ figure ).\footnote{Here, $\tan\gamma < 0.01$ is not covered in our scan, so there is a white strip on the left hand side of these three figures.} In Fig.~\ref{bsg_170180_b_g} at the top left, we see that for the asymmetry $\mathcal{A}^{\text{tot}}_{X_{s}\gamma}$, the parameter space roughly from 0.5\% to 1.5\% is allowed by 3$\sigma$ experiment bound of $\bar{B} \to X_s \gamma$. On the plot of the top right, we can see that the red lines cross a region in where $\Delta\mathcal{A}_{X_{s}\gamma}$ can reach between -1.5\% to -2\%, which would provide a 5$\sigma$ signal at 50 ab$^{-1}$ BELLE-II. However, this asymmetry depends on $\tilde{\Lambda}_{78}$ and is taken the largest allowed value. The bottom untagged asymmetry observable $\mathcal{A}_{\rm CP} (\bar{B} \to X_{s+d} \gamma) $ could achieve up to -3\% which could probe a 5$\sigma$ signal physics beyond the SM and is not affected by $\tilde{\Lambda}_{78}$ which makes it a good testable mode for new physics.

\begin{specialtable}[H]
\caption{BR$(\bar{B} \to X_s \gamma) \,\times 10^{4}$, $\mathcal{A}_{\rm CP} (\bar{B} \to X_s \gamma )$, $\Delta A_{X_s \gamma}$ and $\mathcal{A}_{\rm CP} (\bar{ B} \to X_{s+d} \gamma)$  
for three different values of the scale $\mu_b$ and using LO expression of $C_{7,8}$. Other parameters are fixed as follows:
 $M_{H^{\pm}_2} = 170$ GeV, $M_{H^{\pm}_3} = 180$ GeV, $\theta = - \frac{\pi}{4}$, $\tan\beta = 32$, $\tan \gamma = 1$, $\delta = 2.64$, $m_b = 4.71$ GeV,  $\tilde{\Lambda}^{u}_{17} = -0.66$ GeV, $\tilde{\Lambda}^{c}_{17} = -0.007$ GeV and $\tilde{\Lambda}_{78} = 0.017$ GeV. Reproduced from Ref.~\cite{Akeroyd2020}.}
\label{tab:cpasymmetry_Scale_effect1}
\begin{tabular}{|c|c|c|c|c|}
        \hline
         $\mu_b$ & BR$(\bar{B} \to X_s \gamma)\times 10^{4}$  & $\mathcal{A}_{\rm CP} (\bar{B} \to X_s \gamma )$& $\Delta A_{X_s \gamma}$  & $\mathcal{A}_{\rm CP} (\bar{B} \to X_{s+d} \gamma)$  \\ 
        \hline
        \hline
        $\mu_b= m_b/2$ & 2.912 & -3.170 & -0.111 & -0.974  \\
        $\mu_b = m_b$ & 2.968 & -3.636 & -0.134 &-1.058 \\
        $\mu_b = 2m_b$  & 2.801 & -4.137 & -0.613 & -1.153  \\
        \hline
\end{tabular}
\end{specialtable}

\begin{specialtable}[H]
\caption{BR$(\bar{B} \to X_s \gamma)\,\times 10^{4}$, $\mathcal{A}_{\rm CP} (\bar{B} \to X_s \gamma )$, $\Delta A_{X_s \gamma}$ and $\mathcal{A}_{\rm CP} (\bar{B} \to X_{s+d} \gamma)$  
for three different values of the scale $\mu_b$ and using LO expression of $C_{7,8}$. Other parameters are fixed as follows:
 $M_{H^{\pm}_2} = 170$ GeV, $M_{H^{\pm}_3} = 180$ GeV, $\theta = - \frac{\pi}{4}$, $\tan\beta = 32$, $\tan \gamma = 1$, $\delta = 2.64$, $m_b = 4.77$ GeV, 
 $\tilde{\Lambda}^{u}_{17} = 0$ GeV, $\tilde{\Lambda}^{c}_{17} = 0.0085$ GeV and $\tilde{\Lambda}_{78} = 0.0865$ GeV. Reproduced from Ref.~\cite{Akeroyd2020}.}
\label{tab:cpasymmetry_Scale_effect2}
\begin{tabular}{|c|c|c|c|c|}
        \hline
         $\mu_b$ & BR$(\bar{B} \to X_s \gamma)\times 10^{4}$  & $\mathcal{A}_{\rm CP} (\bar{B} \to X_s \gamma )$ & $\Delta A_{X_s \gamma}$  & $\mathcal{A}_{\rm CP} (\bar{B} \to X_{s+d} \gamma)$   \\ 
        \hline
        \hline
        $\mu_b= m_b/2$ & 2.888 & -1.220 & -0.562 & -1.755  \\
        $\mu_b = m_b$ & 2.931 & -1.663 & -0.673 & -2.151\\
        $\mu_b = 2m_b$  & 2.761 & -2.212 & -0.820 & -2.670 \\
        \hline
\end{tabular}
\end{specialtable}

\begin{specialtable}[H]
\caption{BR$(\bar{B} \to X_s \gamma)\,\times 10^{4}$, $\mathcal{A}_{\rm CP} (\bar{B} \to X_s \gamma )$, $\Delta A_{X_s \gamma}$ and $\mathcal{A}_{\rm CP} (\bar{B} \to X_{s+d} \gamma)$  
for three different values of the scale $\mu_b$ and using LO expression of $C_{7,8}$. Other parameters are fixed as follows:
 $M_{H^{\pm}_2} = 170$ GeV, $M_{H^{\pm}_3} = 180$ GeV, $\theta = - \frac{\pi}{4}$, $\tan\beta = 32$, $\tan \gamma = 1$, $\delta = 2.64$, $m_b = 4.83$ GeV, 
 $\tilde{\Lambda}^{u}_{17} = 0.66$ GeV, $\tilde{\Lambda}^{c}_{17} = 0.010$ GeV and $\tilde{\Lambda}_{78} = 0.19$ GeV. Reproduced from Ref.~\cite{Akeroyd2020}.}
\label{tab:cpasymmetry_Scale_effect3}
\begin{tabular}{|c|c|c|c|c|}
        \hline
         $\mu_b$ & BR$(\bar{B} \to X_s \gamma) \times 10^{4}$  & $\mathcal{A}_{\rm CP} (\bar{B} \to X_s \gamma )$ & $\Delta A_{X_s \gamma}$  & $\mathcal{A}_{\rm CP} (\bar{B} \to X_{s+d} \gamma)$   \\ 
        \hline
        \hline
        $\mu_b= m_b/2$ & 2.865 & 1.145 & -1.223 & -2.123  \\
        $\mu_b = m_b$ & 2.896 & 0.914 & -1.466 & -2.641\\
        $\mu_b = 2m_b$  & 2.724 & 0.581 & -1.785 & -3.323 \\
        \hline
\end{tabular}
\end{specialtable}

We also need to consider the theoretical uncertainty of the asymmetry prediction when varying the energy scale ($\mu_b$) and the hadronic parameter terms. In Tabs.~\ref{tab:cpasymmetry_Scale_effect1}, \ref{tab:cpasymmetry_Scale_effect2}, \ref{tab:cpasymmetry_Scale_effect3}, we have summarized the resulting observables with different chosen energy scales ($\mu_b/2, \mu_b, 2\mu_b$). The pole mass of bottom quark is $4.77\pm0.06$ GeV. In each table, we used the values 4.71 GeV, 4.77 GeV and 4.83 GeV, respectively. 
Different values of NNLO $\bar{B} \to X_s \gamma$ and NLO CP-asymmetries are varied from above three scales. In these tables, it is clear to see the significant effect of varying the hadronic parameter terms in $\mathcal{A}_{\rm CP} (\bar{B} \to X_s \gamma )$, so that even could flip the sign of this observable. Similarly, changing the hadronic parameters also affects the value of  $\Delta A_{X_s \gamma}$ since the contribution to $\tilde{\Lambda}_{78}$ is involved while the effect on $\mathcal{A}_{\rm CP} (\overline B \to X_{s+d} \gamma)$ is less severe. The minimum and the maximum values of the observables  in the three tables are as follows:
\begin{subequations}
\bea
2.724 < &BR( \bar{B} \to X_s \gamma) &  < 2.968 \times 10^{-4} \\
- 4.137< & \mathcal{A}_{\rm CP} (\bar{B} \to X_s \gamma )& < 0.581 \,\,\% \\ 
-1.785 < &\Delta A_{X_s \gamma} & < - 0.111\,\, \%\\ 
- 3.323 < &\mathcal{A}_{\rm CP} (\bar{B} \to X_{s+d} \gamma) & < - 0.974\,\, \%
\eea
\end{subequations}
In conclusion, a full scan over the hadronic parameters $\tilde{\Lambda}^{u}_{17}, \tilde{\Lambda}^{c}_{17}, \tilde{\Lambda}_{78}$ might result in larger asymmetries. The information of BSM parameters will have this uncertainty of scale dependence and if higher order corrections to $\bar{B} \to X_s \gamma$ are included, the scale dependence would decrease.

\subsection{Charged Higgs Bosons in $\bar{B}\to X_s\gamma$ and EDM Constraints} \label{sec: lightmass}

In this section, we combine the limits of the BR($\bar{B}\to X_s\gamma$) and both neutron and electron EDM constraints on the Yukawa coupling mixing parameters plane. We consider three different cases according to hierarchy of the mass of the charged bosons and the top; that is the case where $M_{H^\pm_2}<M_{H^\pm_3}<m_t$, followed by $M_{H^\pm_2}<m_t<M_{H^\pm_3}$, and finally $m_t<M_{H^\pm_2}<M_{H^\pm_3}$. The former two have been carried out with the collider, perturbativity, and top decay width limits together as the description of Appendix~\ref{appendix:b} while the collider search constraints do not provide a strict bound for both heavy masses scenarios (heavier than $ m_t$), so only BR($\bar{B}\to X_s\gamma$) and both neutron and electron EDM constraints are covered for the last case.

\subsubsection{The $M_{H^\pm_2}<M_{H^\pm_3}<m_t$ Case}

\begin{figure}[H]
	\centering
\includegraphics[scale=0.40]{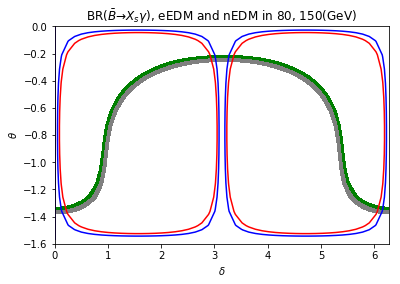}\includegraphics[scale=0.40]{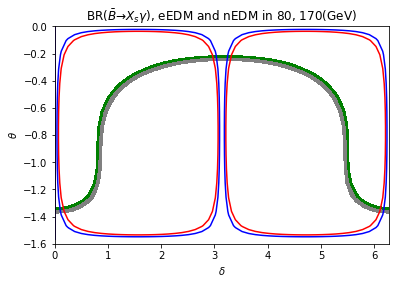}\\
\includegraphics[scale=0.40]{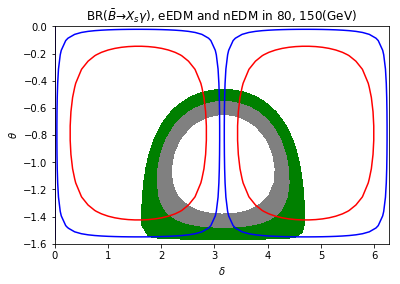}\includegraphics[scale=0.40]{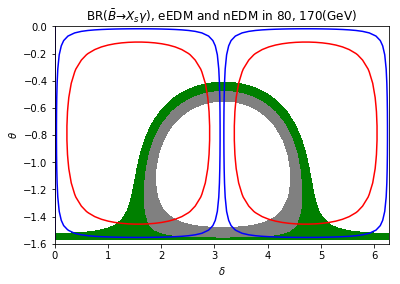}\\
\includegraphics[scale=0.40]{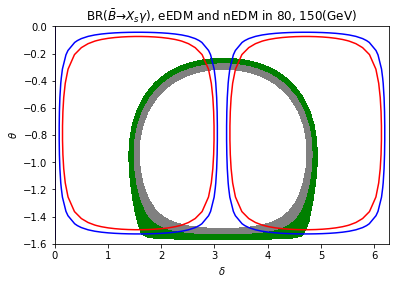}\includegraphics[scale=0.40]{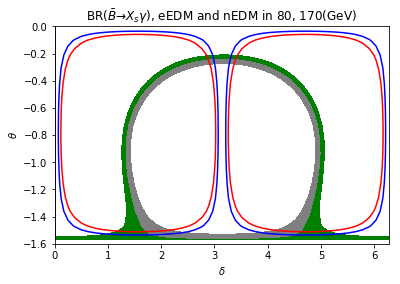}
	\caption{BR($\bar{B} \to X_s  \gamma$) constraint (within the green and grey shaded areas are allowed), eEDM (outside the blue curves are allowed), and nEDM (outside the red curves are allowed) in the [$\delta, \theta$] plane, with $M_{H_2^+} = 80$~GeV and $M_{H_3^+} = 150$ (Left panel) or 170 (Right panel)~GeV.  From top to bottom, $(\tan\beta, \tan \gamma) = (5, 0.5)$; $(5, 1)$; and $(10, 1)$.	Reproduced from Ref.~\cite{Logan:2020mdz}. 
	}
	\label{Fig:edmlow}
\end{figure}

In Fig.~\ref{Fig:edmlow}, we show the constraints on above three limits BR($\bar{B}\to X_s\gamma$), eEDM and nEDM on the $[\delta,\theta]$ plane, with $M_{H_2^+}=80$ GeV and $M_{H_3^+}=150$ GeV ($M_{H_3^+}=170$ GeV) on the left (right), and increasing the value of $(\tan\beta,\tan\gamma)$ in each panel from top to bottom. 
(Notice that the allowed region of $\bar{B}\to X_s\gamma$ is inside the green and grey shaded areas while the allowed region of two EDM constraints is outside the corresponding closed curves.  Our implementation of BR($\bar{B} \to X_s  \gamma$) is given in Appendix~\ref{appendix:a}.) 
We see that we could have a broader interval of $\delta$ around  $\pi$ for large values of $\tan\beta$ and $\tan\gamma$. 
However, if we consider the top-width constraint, we rather need to take lower values of $(\tan\beta,\tan\gamma)$.
There also seems to be a wider band satisfying the $\bar{B}\to X_s\gamma$ constraint on lower value of $M_{H^\pm_3}$. In this case, however,  the $H^\pm\to\tau\nu$ limit is too restrictive on $M_{H^\pm_3}$ as we decrease its mass.

\begin{figure}[H]
\centering
\includegraphics[scale=0.40]{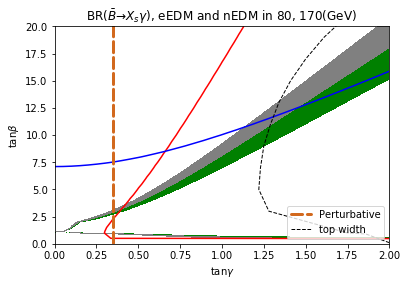}
\includegraphics[scale=0.40]{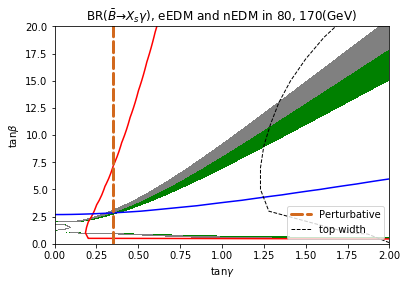}\\
\includegraphics[scale=0.40]{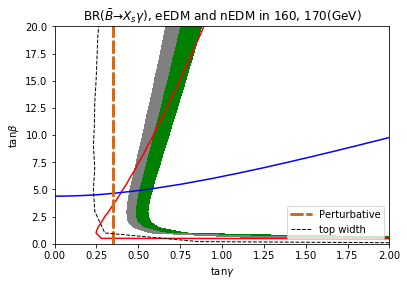}
\includegraphics[scale=0.40]{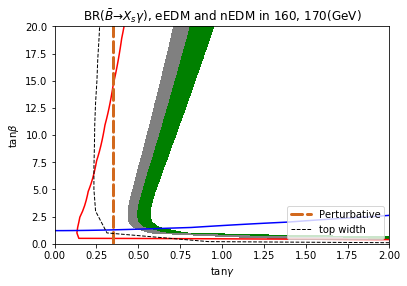}
	\caption{BR($\bar{B} \to X_s  \gamma$) (within the green and grey shaded areas are allowed), eEDM (above the blue line is allowed), and nEDM (the right part of the red line is allowed) in the [$\tan \gamma,\tan \beta$] plane, with $M_{H_3^\pm} = 170$~GeV.  {Left top panel}: $M_{H_2^\pm} = 80$~GeV, $\theta = -0.3$, $\delta = 0.96\pi$. {Right top panel}: $M_{H_2^\pm} = 80$~GeV, $\theta = -0.3$, $\delta = 0.985\pi$. {Left bottom panel}: $M_{H_2^\pm} = 160$~GeV, $\theta = -0.5$, $\delta = 0.8\pi$. {Right bottom panel}: $M_{H_2^\pm} = 160$~GeV, $\theta = -0.5$, $\delta = 0.95\pi$.
The top-quark width (black dotted line) and perturbativity (orange dashed line), wherein the allowed region is to the right of the respective curves, are presented in these four figures.
These constraints are listed in Appendix~\ref{appendix:b} and reproduced from Ref.~\cite{Logan:2020mdz}. 
	}
	\label{Fig:lowmass-summary}
\end{figure}

In the upper panel of Fig.~\ref{Fig:lowmass-summary}, we show the case $M_{H^\pm_2}=80$~GeV and  $M_{H^\pm_3}=170$~GeV with ($\theta$, $\delta$) values as explained in the caption. The black dotted line is the top width limit, the right side of the bound are the allowed region. 
The top-quark width measurement strongly constrains this case, excluding $\tan\gamma$ with very low values.
In the lower panel, $M_{H^\pm_2}=160$~GeV and $M_{H^\pm_3}=170$~GeV are presented.
Here, the width of the top quark is less constraining than the former one, and $\tan\gamma$ could have very low values.  
A more extensive range of the CP-violating phase $\delta$ is allowed when two states are closer to the degenerate mass case.

\subsubsection{The $M_{H_2^\pm}<m_t<M_{H_3^\pm}$ Case}

In Fig.~\ref{Fig:80,200EDM}, the constraints on $\bar{B}\to X_s\gamma$, eEDM and nEDM within the $[\delta,\theta]$ plane for $M_{H_2^\pm} = 80$~GeV, $M_{H_3^\pm} = 200$~GeV, and $\tan\beta =5\,(10)$ left\,(right) and $\tan\gamma =1$ are presented.
From these plots, the CP-violation phase $\delta$ has to be very close to $\delta = n \pi$  to satisfy all three constraints at once and such scenario is forced that the solutions have to be very close to the CP-conserving limit. In addition,$\bar B \to X_s \gamma$ constraint tends to prefer $\delta \simeq \pi$.
\begin{figure}[H]
	\centering
\includegraphics[scale=0.40]{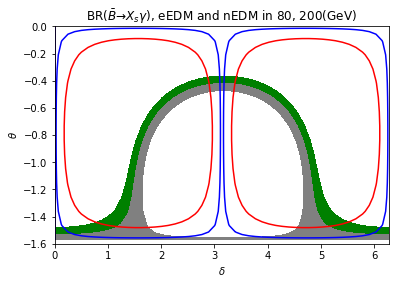}
\includegraphics[scale=0.40]{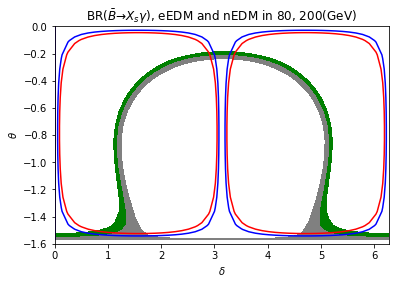}
	\caption{BR($\bar{B} \to X_s  \gamma$) (within the green and grey shaded areas are allowed), eEDM (outside the blue curves are allowed), and nEDM (outside the red curves are allowed) in the [$\delta, \theta$] plane, with $M_{H_2^\pm} = 80$~GeV, $M_{H_3^\pm} = 200$~GeV. {Left panel:} $\tan\gamma = 1$, and $\tan\beta = 5$. {Right panel:}  $\tan\gamma = 1$, and $\tan\beta = 10$.  Reproduced from Ref.~\cite{Logan:2020mdz}. 
	}
	\label{Fig:80,200EDM}
\end{figure}

\begin{figure}[H]
\centering
\includegraphics[scale=0.40]{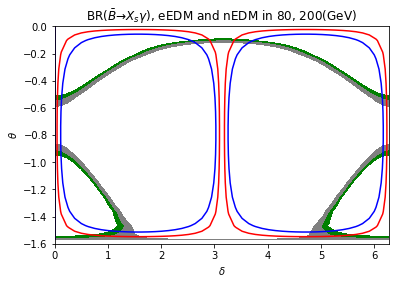}
\includegraphics[scale=0.40]{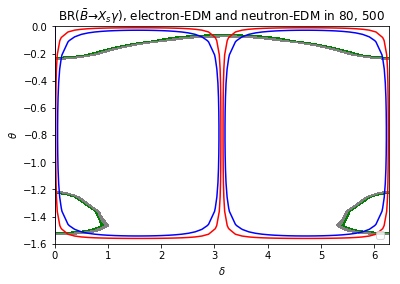}\\
\includegraphics[scale=0.40]{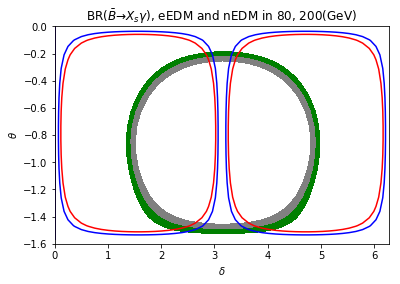}
\includegraphics[scale=0.40]{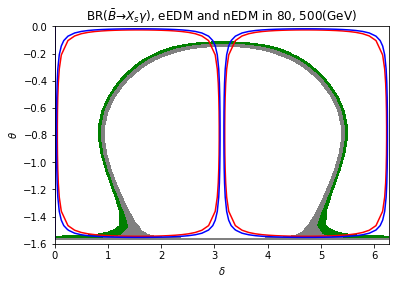}
\caption{BR($\bar{B} \to X_s  \gamma$) (within the green and grey shaded areas are allowed), eEDM (outside the blue curves are allowed), and nEDM (outside the red curves are allowed) in the [$\delta, \theta$] plane, with $M_{H_2^\pm} = 80$~GeV and $\tan\beta = 20$.  {Left top panel:} $M_{H_3^\pm} = 200$~GeV, $\tan\gamma = 1$. {Right top panel:} $M_{H_3^\pm} = 500$~GeV, $\tan\gamma = 1$. {Left bottom panel:} $M_{H_3^\pm} = 200$~GeV, $\tan\gamma = 2$. {Right bottom panel:} $M_{H_3^\pm} = 500$~GeV, $\tan\gamma = 2$. Reproduced from Ref.~\cite{Logan:2020mdz}. 
}
\label{fig:deltatheta20}
\end{figure}

In Fig.~\ref{fig:deltatheta20}, we show the effect of variation in $\tan\gamma$ with same $\tan\beta =20$ and two different $M_{H_3^\pm}$ states while keeping $M_{H_2^\pm} = 80$~GeV.
It is difficult to find surviving regions that satisfy all three constraints when $M_{H_3^\pm}$ increases from 200 to 500 GeV. In comparison with the Fig.~\ref{Fig:80,200EDM}, we also see that larger $\tan\beta$ leads to tighter constraints on the nEDM while tighter eEDM constraints require smaller $\tan\gamma$ values.

In Fig.~\ref{Fig:80,200constraints}, we show the same plane according to $[\tan\gamma,\tan\beta]$ and two values of $\delta$ very close to $\delta = \pi$. In this figure, the constraints on the top-quark width and perturbativity of the $H_i^+ b\bar t$ vertex which is given in Appendix~\ref{appendix:b} are also plugged in. For all the parameter region, the collider limits ($H^{\pm} \to cb/ cs/ \tau \nu_\tau$) are satisfied.
In the case of $\tan\gamma > 1.5$ and $\tan\beta > 8$, we can satisfy all other constraints, but the CP-phase $\delta$ should roughly be close to $\pi$ which is at least $\delta = 0.975\pi$ to $0.985\pi$ for, e.g., $\theta=-0.3$.  
\begin{figure}[H]
\centering
\includegraphics[scale=0.40]{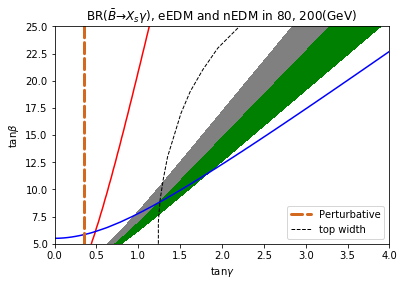}
\includegraphics[scale=0.40]{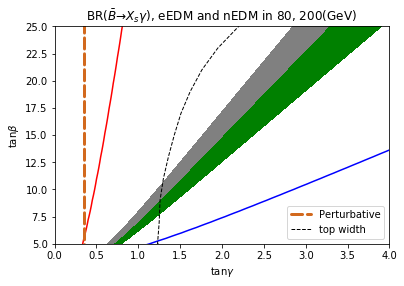}
	\caption{BR$\bar{B} \to X_s \gamma$ (regions within the green and grey shaded areas are allowed), eEDM (above the blue line is allowed region) and nEDM ( right part of the red line is allowed) in the [$\tan \gamma,\tan \beta$] plane, with $M_{H_2^\pm} = 80$~GeV, $M_{H_3^\pm} = 200$~GeV, $\theta = -0.3$. {Left panel:}  $\delta = 0.975\pi$. {Right panel:}  $\delta =0.985\pi$.  The top-quark width (black dotted line) and perturbativity (orange dashed line) constrains are also implemented. The region to the right of these two lines is allowed. Reproduced from Ref.~\cite{Logan:2020mdz}. }
	\label{Fig:80,200constraints}
\end{figure}

\subsubsection{The $m_t<M_{H^\pm_2}<M_{H^\pm_3}$ Case}

\begin{figure}[H]
	\centering
	\includegraphics[scale=0.40]{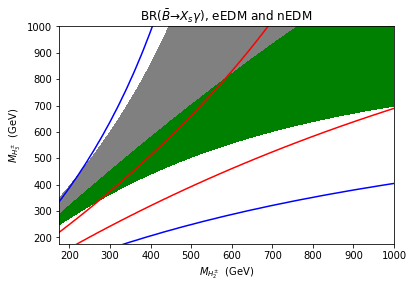}	\includegraphics[scale=0.40]{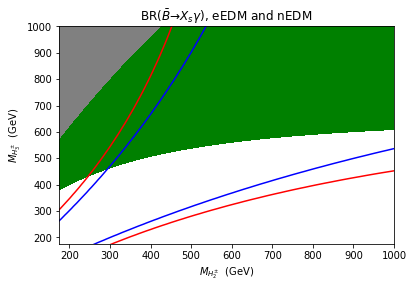}\\	\includegraphics[scale=0.40]{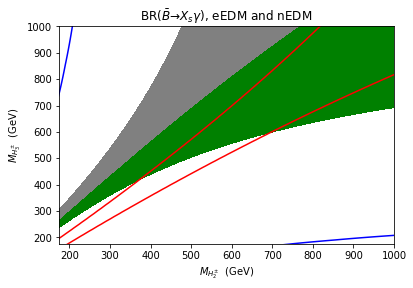}	\includegraphics[scale=0.40]{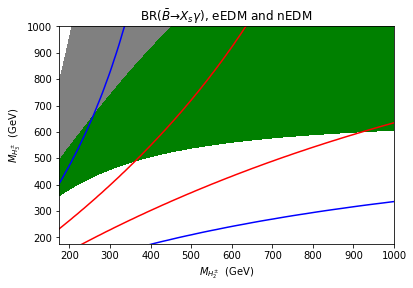}
	\caption{BR($\bar{B} \to X_s  \gamma$) (regions in the green and grey shaded areas are allowed),  eEDM (regions between the blue lines are allowed), and  nEDM (regions between the red lines are allowed) in the [$M_{H^{\pm}_2} , M_{H^{\pm}_3} $] plane, for 
 $\theta = -0.476\pi$ and $\delta = 0.5\pi$ (i.e., maximal CP-violation). {Left top panel}: $\tan\beta = 20, \tan\gamma = 1$. {Right top panel}: $\tan\beta = 20, \tan\gamma = 2$. {Left bottom panel}: $\tan\beta = 40, \tan\gamma = 1$. {Right bottom panel}: $\tan\beta = 40, \tan\gamma = 2$. Reproduced from Ref.~\cite{Logan:2020mdz}. 
}
\label{fig:theta2105}
\end{figure}

In the case that two charged Higgs masses are heavier than $m_t$, the parameter space is no longer limited strictly to collider searches since the production of charged Higgs leads small cross section, so only the BR($\bar{B} \to X_s  \gamma$), eEDM and nEDM constraints on the [$M_{H^{\pm}_2},M_{H^{\pm}_3}$] plane with different choices for the mixing parameters ($\tan\beta$, $\tan \gamma$, $\theta$, and $\delta$) will be presented. 
In Figure \ref{fig:theta2105}, the nEDM will be the most constraining limit of all above three limits which has shown in two bottom panels when $\tan\beta = 40$. In contrast, with the choice of $\tan\gamma =2$ while $\tan\beta = 20$ on the right top panel of Figure \ref{fig:theta2105}, the eEDM becomes the most constraining limit. The EDM constraint provides a funnel shape of the charged Higgs masses plane.

\begin{figure}[H]
	\centering
	\includegraphics[scale=0.4]{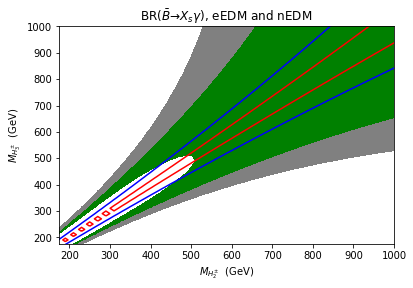}	\includegraphics[scale=0.4]{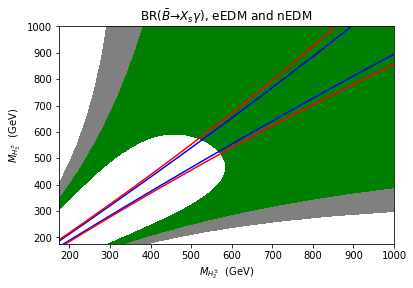}\\	\includegraphics[scale=0.4]{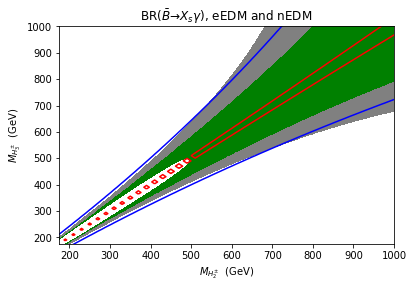}	\includegraphics[scale=0.4]{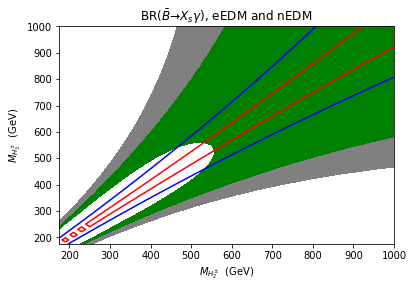}
	\caption{Same mass plane as Fig.~\ref{fig:theta2105} and $\theta=-\pi/4$. Reproduced from Ref.~\cite{Logan:2020mdz}. 
}
\label{fig:theta405}
\end{figure}

We present in Fig.~\ref{fig:theta405} the case of $\theta = -\pi/4$, $\delta = 0.5\pi$ (maximum CP-violation) and the rest of the parameters as in Fig.~\ref{fig:theta2105}. For this value of $\theta$, the eEDM and nEDM made a funnel shape of the mass diagonal, illustrating a GIM-like cancellation mechanism \cite{Glashow1970}, which suppress flavour changing neutral currents through induced loop diagrams by summation of multiple flavour of quarks. The summation of charged Higgs states in this case provide the cancellation of CP-violation and mass degeneracy of charged scalars masses make the cancellation exact.
However, in the case of $\delta = 0.5\pi$, it is intriguing that even the exact degeneracy case between the two charged Higgs states $H_2^\pm$ and $H_3^\pm$ fails the $\bar B\to X_s\gamma$ for charged Higgs masses below 500 GeV. 
This funnel-like shape region for EDMs is a general feature independent of the value of $\tan\beta$, as long as $\tan\gamma$ remains small.

\begin{figure}[H]
	\centering
	\includegraphics[scale=0.4]{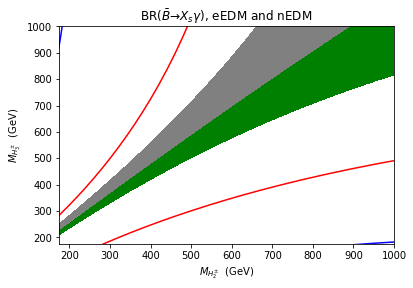}	\includegraphics[scale=0.4]{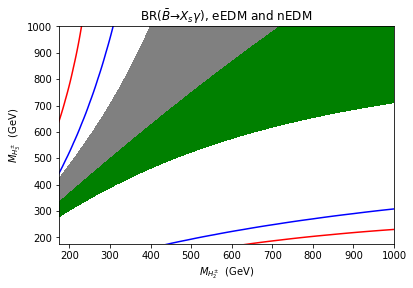}\\	\includegraphics[scale=0.4]{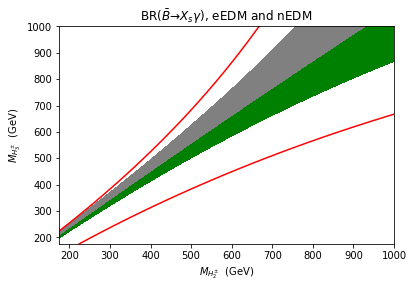}	\includegraphics[scale=0.4]{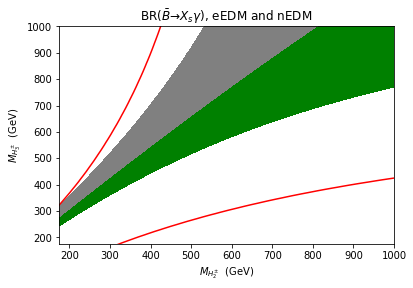}
	\caption{Same mass plane as Fig.~\ref{fig:theta2105} and $\delta=0.85\pi$. Reproduced from Ref.~\cite{Logan:2020mdz}. 
}
\label{fig:theta2185}
\end{figure}

\begin{figure}[H]
	\centering
	\includegraphics[scale=0.4]{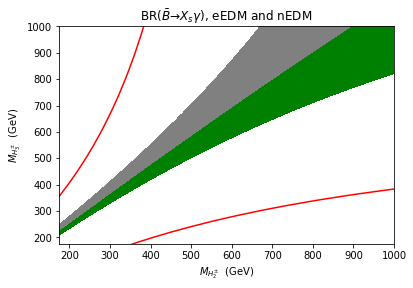}	\includegraphics[scale=0.4]{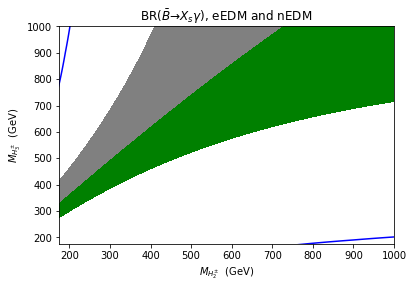}\\	\includegraphics[scale=0.4]{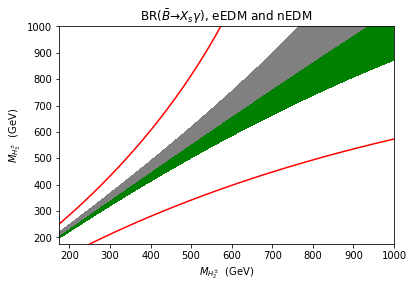}	\includegraphics[scale=0.4]{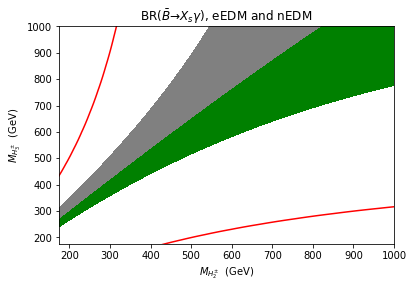}
	\caption{Same mass plane as Fig.~\ref{fig:theta2105} and $\delta=0.9\pi$. Reproduced from Ref.~\cite{Logan:2020mdz}. 
}
\label{fig:theta219}
\end{figure}

{In Figs.~\ref{fig:theta2185} and \ref{fig:theta219}, the same parameters as in Fig.~\ref{fig:theta2105} are used but with $\delta = 0.85\pi$ and $\delta = 0.9\pi$, respectively, i.e., we show two cases  close to the CP-conserving limit. A large expense of parameter space is allowed by both the eEDM and nEDM. In fact, the EDM constraints no longer limit strictly the parameter space even for masses below 500 GeV. In the top left of Figure \ref{fig:theta2185}, only the nEDM bound (the region between the red lines) is presented while the eEDM (blue lines) constraints are satisfied by almost all the parameter space. Similarly, other subplots of Figs.~\ref{fig:theta2185} and \ref{fig:theta219} show that the parameter space is constrained severely by $\bar{B} \to X_s \gamma$ instead. }

\section{Discussion}\label{section4}

In the CP-asymmetry section, we have studied the dependence on the two charged Higgs masses as well as the mixing parameters for a benchmark point that would survive for large values of three CP-asymmetry observables $\mathcal{A}^{\text{tot}}_{X_{s}\gamma}$, $\mathcal{A}_{\rm CP} (\overline B \to X_{s+d} \gamma) $ and $\Delta\mathcal{A}_{X_{s}\gamma}$. On the one hand, both observables $\mathcal{A}^{\text{tot}}_{X_{s}\gamma}$ and $\Delta\mathcal{A}_{X_{s}\gamma}$ are strongly correlated to the non-perturbative hadronic parameters,  Eqs.~(\ref{acpshortlong}) and (\ref{eq:delta_cp}), such that the latter could affect the contribution from the two charged Higgs boson states. On the other hand, $\mathcal{A}_{\rm CP} (\overline B \to X_{s+d} \gamma) $ is independent of the hadronic parameters. The most remarkable feature of these results is seen in the bottom panel of Fig.~\ref{bsg_170180_b_g}, where $\mathcal{A}_{\rm CP} (\overline B \to X_{s+d} \gamma)$ could achieve values larger than 2.5\%, which could then to probe a 5$\sigma$ signal to BSM. Thus, such an observable without uncertain parameters, $\mathcal{A}_{\rm CP} (\overline B \to X_{s+d} \gamma)$, becomes a hallmark signal of a CP-violating 3HDM. One more relevant point is that 3HDM would have more parameter space for the large asymmetries than in the A2HDM~\cite{Jung:2012vu,Jung2014}. In fact, the A2HDM contains only one charged scalar state so that there is less possibility for cancellation in the corresponding contributions to $\bar{B} \to X_s \gamma$ compared to the 3HDM (also, there are no additional $X_3$, $Y_3$ couplings as in the 3HDM).

We have seen in Figures~\ref{fig:theta2185} and \ref{fig:theta219} that the effects of the CP-violating phase in processes with a virtual exchange of charged scalars can be arbitrarily suppressed when the two charged bosons are nearly degenerate in mass, avoiding then the EDM constraints. 
Such phenomenon can be understood as an equivalent of the GIM mechanism and the allowed region along the EDM constraints shows a funnel shape is again presenting the impact on GIM-like cancellation mechanism generated by charged Higgs mass degeneracy.
More specifically, the degeneracy of $H^{\pm}_{2}$ and $H^{\pm}_{3}$ affects the CP-violating phase and the mixing angle ($\delta$ and $\theta$) to be non-physical. The fermion line connection with the internal charged Higgs propagator brings one factor of $X_i^*$, $Y_i^*$ or $Z_i^*$ on one side and one factor of $X_i$, $Y_i$ or $Z_i$ on the other side.  The combinations $X_iX_i^*$, $Y_iY_i^*$, and $Z_iZ_i^*$ are purely real and would not lead any CP-odd terms, and then only the combinations $X_iY_i^*$, $X_iZ_i^*$ and $Y_iZ_i^*$ (or their complex conjugates) are correlated to the CP-odd observables. For instance, the terms like $X_iY_i^*$ in rotation matrix of charged Higgs in the Democratic 3HDM can be expressed as:
\bea
X_i Y_i^* = - \frac{U^{\dagger}_{1i} U_{i2}}{U^{\dagger}_{11} U_{12}},
\eea
where $i = 2$ or 3. In this case, $U_{1j} = v_j/v$ affects the denominator to be real under construction. The contributing diagrams would take the sum of two charged states which is involved in the computation of the CP-odd observables in this context, yielding
\begin{equation}
	\sum_{i = 2}^3 {\rm Im} (X_i Y_i^*) f(M_{H^+_i})
	= - \frac{1}{U_{11}^{\dagger} U_{12}} \left[ {\rm Im} (U_{12}^{\dagger} U_{22}) f(M_{H^\pm_2})
		+ {\rm Im} (U_{13}^{\dagger} U_{32}) f(M_{H^\pm_3}) \right],
	\label{eq:gimmech1}
\end{equation}
where $f(M_{H^\pm_i})$ is relate to the diagram on the specific charged Higgs boson mass. In the limit of mass degeneracy scenario which gives $M_{H^\pm_2} =M_{H^\pm_3} = m$, we could have:
\begin{equation}
	\sum_{i = 2}^3 {\rm Im} (X_i Y_i^*) f(m) 
	=  - \frac{1}{U_{11}^{\dagger} U_{12}} {\rm Im} \left[ \sum_{i = 1}^3 U_{1i}^{\dagger} U_{i2} \right] f(m) 
	= - \frac{1}{U_{11}^{\dagger} U_{12}} {\rm Im} (\delta_{12}) f(m) = 0,
\end{equation}
where $\delta_{12}$ is the $(1,2)$ element of the Kronecker delta. In fact, this also shows ${\rm Im}(X_2 Y_2^*) = - {\rm Im}(X_3 Y_3^*)$ due to the unitary charged Higgs mixing matrix Eq.~(\ref{eq:Uexplicit}). Similarly, the contribution to eEDM, which requires imaginary parts of $Y_iZ_i^*$, will also have the same appearance. With a small non-zero mass splitting $\Delta M_{H^\pm} \ll M_{H^\pm}$, CP-violating amplitudes must be linear corresponding to $\Delta M_{H^\pm} /M_{H^\pm}$, where $\Delta M_{H^\pm} \equiv M_{H^\pm_3} - M_{H^\pm_2}$ and $M_{H^\pm} \equiv (M_{H^\pm_3} + M_{H^\pm_2})/2$. Thus, the typical CP-violating  contributions to the Democratic 3HDM give an interesting contributions to both the eEDM and nEDM. For other types of 3HDM, however, CP-violation is more limited, due to the combination of Yukawa couplings $X_i$, $Y_i$, and $Z_i$. In specific, there will be no contribution to ${\rm Im}(-Y_i^* Z_i)$ to the eEDM in Type-I and Type-Y (Flipped) 3HDMs because in those models $Y_i = Z_i$. On the one hand, Type-I and Type-X (Lepton-specific) 3HDMs would not perform any CP-violation in nEDM from charged Higgs contribution since $X_i = Y_i$ in these models.  On the other hand, such a GIM-like cancellation mechanism does not appear in the $\bar{B} \to X_s \gamma$ constraint. The reason is that this observable covers both real and imaginary contributions from $X_iY_i^*$ terms and the real $X_2Y^*_2$ and$X_3Y^*_3$ components do not have strongly correlation as their imaginary parts. Thus, the shape of $\bar{B} \to X_s \gamma$ is different than the EDMs and the available solution will be a judicious choice of $\theta$ with certain values of $\tan\beta$ and $\tan\gamma$ (compare to Figs.~\ref{fig:theta2105} and \ref{fig:theta405}). 

In the end, the surviving parameter region provides favourable value of CP-phase $\delta$ to be close to $\pi$ from most three constrained limits, $\bar{B} \to X_s \gamma$ and n(e)EDMs. The preference region can be seen from Fig \ref{Fig:80,200EDM} to \ref{fig:deltatheta20} and lying near the middle of $2\pi$ values on x-axis.  $\tan\beta$ is favoured with large values and $\tan\gamma$ will have to be small in contrast. For masses of two charged Higgs states, two states could be smaller than top quark but the CP-phase actually limits the available mass values.
\section{Conclusions}
\label{section5}
Our analysis concludes that the CP-asymmetry observables from the 3HDM could be a potential new physics experimental discovery mode of such a BSM scenario . 
Also, in the case of mass degeneracy in the charged Higgs boson sector , a GIM-like cancellation mechanism allows one avoiding EDM constraints , both neutron and electron ones . 
In future research , it would be interesting to analyse the possibility of CP-violating phases from both the neutral and charged Higgs sectors. Our studies can help expand the understanding of CP-violation of the extended scalar sectors and spark further interest in MHDMs .



\vspace{6pt} 



\authorcontributions{Conceptualization, A.A., S.M., H.L. and M.S.; methodology, A.A. and S.M; software, D.R.-C.,T.S. and M.S.; validation, D.R.-C.,T.S. and M.S.; formal analysis, A.A.,S.M. and H.L.; investigation, D.R.-C. and M.S.; resources, A.A., H.L. and T.S.; data curation, D.R.-C.,T.S. and M.S.; writing---original draft preparation, A.A., H.L., D.R.-C. and M.S.; writing---review and editing, S.M. ; visualization, D.R.-C. and M.S.; supervision, A.A.,S.M.,H.L. and T.S. All authors have read and agreed to the published version of the manuscript.}

\funding{AA and SM are supported in part through the STFC Consolidated Grant ST/L000296/1.
SM is supported in part also through the NExT Institute.  TS is supported in part by JSPS
KAKENHI Grant Number 20H00160. TS and SM are partially supported by the Kogakuin
University Grant for the project research ”Phenomenological study of new physics models
with extended Higgs sector. H.E.L. iso supported by the Natural Sciences and Engineering Research
Council of Canada. D.R.-C. is supported by the Royal Society Newton International Fellowship NIF/R1/180813 and by the National Science Centre (Poland)
under the research Grant No. 2017/26/E/ST2/00470.  All authors acknowledge the H2020-
MSCA-RISE-2014 grant no. 645722 (NonMinimalHiggs) for travel funding.}

\acknowledgments
{D.R.-C. and M.S. thank Carleton
University for hospitality during the initial stages of this work. M.S. thanks Peking University for hospitality during the final stages of this work. The authors thank Shinya
Kanemura and Kei Yagyu for useful conversations.}

\conflictsofinterest{The authors declare no conflict of interest.} 





\appendixtitles{yes} 
\appendixstart
\appendix
\section{The Charged Higgs Boson Yukawa couplings}
\label{appendix:char}

In this section,  the combinations of the Yukawa coupling coefficients $X_i$, $Y_i$, and $Z_i$   ($i = 2,3$) are presented, and illustrate the correlation relationship between couplings and the four mixing parameters ($\theta$, $\tan\gamma$, $\tan\beta$ and $\delta$) in the Democratic 3HDM.  The shorthand notation $s_\theta$, $c_\theta$, $t_\theta$ are $\sin\theta$, 
$\cos\theta$, and $\tan\theta$, respectively, and analogously for the other mixing parameter angles.

From Eq.~(\ref{eq:Uexplicit}) and (\ref{XYZ}), the Yukawa coupling coefficients in our parametrization are as follows:
\bea
X_2 &=&  \frac{U^\dagger_{12}}{U^\dagger_{11}} =   \frac{- c_\theta s_\beta (c_\delta + i s_\delta ) -  s_\theta c_\gamma c_\beta}{c_\beta s_\gamma},\\
Y_2  &=& - \frac{U^\dagger_{22}}{U^\dagger_{21}} =  \frac{- c_\theta c_\beta (c_\delta + i s_\delta) +  s_\theta c_\gamma s_\beta}{s_\beta s_\gamma},\\
Z_2 &=& \frac{U^{\dagger}_{32}}{U^\dagger_{31}} = \frac{s_\theta s_\gamma}{c_\gamma}, \\
X_3 &=& \frac{U^\dagger_{13}}{U^\dagger_{11}} = \frac{s_\theta s_\beta  (c_\delta + i s_\delta)  - c_\theta c_\gamma  c_\beta }{c_\beta s_\gamma} ,\\
Y_3  &=& - \frac{U^\dagger_{23}}{U^\dagger_{21}} =  \frac{   s_\theta c_\beta (c_\delta + i s_\delta) +  c_\theta c_\gamma s_\beta}{s_\beta s_\gamma}, \\
Z_3 &=& \frac{U^{\dagger}_{33}}{U^\dagger_{31}} = \frac{c_\theta s_\gamma}{c_\gamma}.
\eea

The Yukawa combinations that contribute EDM calculations are:
\bea
{\rm Im}(- X_2Y_2^* ) &=& \frac{s_\theta c_\theta s_\delta}{s_\beta c_\beta s_\gamma t_\gamma} 
	= - {\rm Im}(- X_3 Y_3^*) , \\
{\rm Im}(- Y_2^* Z_2) &=& -\frac{s_\theta c_\theta s_\delta}{t_\beta c_\gamma}
	= - {\rm Im}(- Y_3^* Z_3).
\eea

The following relations contribute to BR($\bar B \to X_s \gamma$) calculation.
The real components of $X_iY_i^*$ ($i = 2,3$) are as follows:
\bea
{\rm Re}(X_2Y_2^*) 
&=&\frac{c_\theta^2}{s_\gamma^2} + \frac{c_\delta c_\theta s_\theta}{t_\beta t_\gamma s_\gamma} - \frac{c_\delta t_\beta c_\theta s_\theta }{t_\gamma s_\gamma} - \frac{s_\theta^2}{t_\gamma^2}, \\
{\rm Re}(X_3Y_3^*) 
&=&  \frac{s_\theta^2}{s_\gamma^2} + \frac{c_\delta t_\beta c_\theta s_\theta }{t_\gamma s_\gamma}  - \frac{c_\delta c_\theta s_\theta }{t_\beta t_\gamma s_\gamma} - \frac{c_\theta^2}{t_\gamma^2}.
\eea
Finally for $|Y_2^2|$ and $|Y_3^2|$ we have:
\bea
|Y_2^2| 
&=& \frac{c_\delta^2 c_\theta^2}{t_\beta^2 s_\gamma^2} - \frac{s_\delta^2 c_\theta^2}{t_\beta^2 s_\gamma^2} - \frac{2 c_\delta c_\theta s_\theta}{t_\beta t_\gamma s_\gamma} + \frac{s_\theta^2}{t_\gamma^2}, \\
|Y_3^2| 
&=& \frac{c_\delta^2 s_\theta^2}{t_\beta^2 s_\gamma^2}  - \frac{s_\delta^2 s_\theta^2}{t_\beta^2 s_\gamma^2} + \frac{2 c_\delta c_\theta s_\theta}{t_\beta t_\gamma s_\gamma} + \frac{c_\theta^2}{t_\gamma^2}. 
\eea

\section{$\bar{B} \to X_s \gamma$ and Charged Higgs Boson Yukawa Couplings} \label{appendix:a}

\subsection{Input Parameters for $\bar{B} \to X_s  \gamma$ and CP-asymmetries}

Our input SM parameters for BR$(\bar{B} \to X_s \gamma)$ and CP-asymmetries numerical evaluations are taken from Tab.~\ref{tab:bsg}. In such a case, the pole masses of the charm-, bottom- and top-quark are $m_c$, $m_b$ and $m_t$, respectively, and $A$, $\lambda$, $\bar{\rho}$ and $\bar{\eta}$ are the Wolfenstein parameters of the CKM matrix taken from Ref. \cite{Zyla2020}. 

\begin{specialtable}[H]
	\begin{center}
\hspace*{-0.6cm}
\caption{Input SM parameters. The central value of $\bar{B}\to X_s \gamma$ is calculated through these inputs. 
We follow \cite{Borzumati1998} for fermion mass choices. 
The Wolfenstein parameters of the CKM matrix are taken from Ref. \cite{Zyla2020}.  }
\begin{tabular}{ccc}
	\toprule
	$m_c/ m_b$ = 0.29 & $m_b - m_c$ = 3.39 GeV & $m_t$ = 173 GeV  \\	
	$\alpha_{\text{em}}$ = 1/ 130.3 & $M_Z$ = 91.1875 GeV & $M_{W^\pm}$ = 80.33 GeV   \\
	$\lambda$ = 0.22650 &$A$ = 0.790 & $\bar{\rho}$ = 0.141   \\
	$\bar{\eta}$ = 0.357 & $G_{\text{F}}$ = 1.1663787 $\times 10^{-5}$ GeV$^{-2}$ &  $\alpha({M_Z})$ = 0.119 \\  BR$_{SL}$ = 0.1049 & & \\
	\bottomrule
\end{tabular}
\label{tab:bsg}
	\end{center}
\end{specialtable}

\subsection{Experimental and Theoretical Evaluation of $\bar{B} \to X_s  \gamma$}

Recent values of the average experimental measurement \cite{Amhis2019} of BR($\bar{B} \to X_s \gamma$) is: 
\bea
{\rm BR}(\bar{B} \to X_s \gamma)^{\rm exp} = (3.32 \pm 0.15) \times 10^{-4} \,\, \text{with} \,\,\, E_{\gamma} > 1.6\, \text{GeV},
\label{eq:bsgexp} 
\eea
The NNLO SM prediction \cite{Misiak2020} for the BR($\bar{B} \to X_s \gamma$) is as follows:
\bea
{\rm BR}(\bar{B} \to X_s \gamma)^{\rm SM} = (3.40 \pm 0.17) \times 10^{-4} \,\, \text{with} \,\,\, E_{\gamma} > 1.6\, \text{GeV}.
\label{eq:SMbsg}
\eea
The BR$(\bar{B} \to X_s \gamma)$ numerical evaluation in our analysis is taken from the computed 2HDM explicit formulas with leading order (LO) and next-to-LO (NLO) effective Wilson coefficients, which performs from the larger $\mu_W$ scale to lower $\mu_b$ with the scheme adopted in \cite{Borzumati1998} and extrapolated it to 3HDM as in Ref. \cite{Akeroyd2020}. 
We implemented the above EDM calculation in a Python programme to obtain the results discussed in this work.

The two effective LO Wilson coefficients Charged Higgs contributions at the $\mu_W$ scale are $C^{0,\text{eff}}_7 (\mu_W)$ and $C^{0,\text{eff}}_8 (\mu_W)$, obtained as follows:
\bea \label{eq:c07}
C^{0,\text{eff}}_7 (\mu_W) &=& C^{0}_{7,SM}  + |Y_2 |^2 C^{0}_{7,Y_2Y_2} + |Y_3 |^2 C^{0}_{7,Y_3Y_3}  \nonumber \\ && + (X_2Y_2^*) C^{0}_{7,X_2Y_2} + (X_3Y_3^*) C^{0}_{7,X_3Y_3}, \\
C^{0,\text{eff}}_8 (\mu_W) &=& C^{0}_{8,SM}  + |Y_2 |^2 C^{0}_{8,Y_2Y_2} + |Y_3 |^2 C^{0}_{8,Y_3Y_3}  \nonumber \\  && + (X_2Y_2^*) C^{0}_{8,X_2Y_2} + (X_3Y_3^*) C^{0}_{8,X_3Y_3} ,
\eea
where $|Y_2 |^2$, $|Y_3 |^2$, $(X_2Y_2^*)$, and $(X_3Y_3^*)$ are the combination of charged Higgs mixing couplings.  The LO Wilson coefficients are $C^{0,\text{eff}}_i (\mu_W) = 0$ ($i =1,3,4,5,6$) while $C^{0,\text{eff}}_2 (\mu_W)= 1$. The BSM charged Higgs contributions are functions of $m^2_t / M^2_{H^{\pm}_2}$ and $m^2_t / M^2_{H^{\pm}_3}$ and the SM contribution are functions of $m^2_t / M^2_{W}$. LO Wilson coefficients $C^{0}_{j,X_2Y_2}, C^{0}_{j,X_3Y_3} \,\, (j = 7,8), C^{0}_{i,{\rm SM}}$ will absorb these mass squared ratio terms.

At the matching scale ($\mu_{W}$), the NLO Wilson coefficients for calculation are:
\bea
C^{1,\text{eff}}_1 (\mu_W) &=&  15 + 6 \hspace{0.2cm}  \text{ln} \frac{\mu^2_{W}}{M^2_W} ,\\
C^{1,\text{eff}}_4 (\mu_W) &= & E_0 + \frac{2}{3}  \hspace{0.2cm}  \text{ln}  \frac{\mu^2_{W}}{M^2_W} + |Y_2 |^2 E_{H_2} + |Y_3 |^2 E_{H_3}, \\
C^{1,\text{eff}}_i (\mu_W) &=& 0 \quad (i = 2,3,5,6), \\
C^{1,\text{eff}}_7 (\mu_W) &=& C^{1,\text{eff}}_{7,SM}(\mu_W)  + |Y_2 |^2 C^{1,\text{eff}}_{7,Y_2Y_2}(\mu_W)  + |Y_3 |^2 C^{1,\text{eff}}_{7,Y_3Y_3}(\mu_W)  \nonumber \\ & & + (X_2Y_2^*) C^{1,\text{eff}}_{7,X_2Y_2}(\mu_W) + (X_3Y_3^*) C^{1,\text{eff}}_{7,X_3Y_3}(\mu_W), \\
C^{1,\text{eff}}_8 (\mu_W) &=& C^{1,\text{eff}}_{8,SM}(\mu_W)  + |Y_2 |^2 C^{1,\text{eff}}_{8,Y_2Y_2}(\mu_W)  + |Y_3 |^2 C^{1,\text{eff}}_{8,Y_3Y_3}(\mu_W) \nonumber \\ & & + (X_2Y_2^*) C^{1,\text{eff}}_{8,X_2Y_2}(\mu_W) + (X_3Y_3^*) C^{1,\text{eff}}_{8,X_3Y_3}(\mu_W).
\eea
All Explicit forms can be found in Ref.~\cite{Borzumati1998}. In order to compose to scale at $\mu=m_b$, we evaluate the Wilson coefficients from scale $\mu = M_W$ by running the renormalization group. The decay rate of $\bar{B} \to X_{s} \gamma $  is then evaluated through:
\bea
\Gamma (\bar{B} \to X_{s} \gamma) &=& \frac{G^2_F}{32\pi^4} |V^{*}_{ts} V_{tb} |^2 \alpha_{em} m^5_b \nonumber   \\ &\times &
\Bigg \{ |\bar{D} |^2 + A + \frac{\delta^{NP}_{\gamma}}{m^2_b} |\text{C}^{0,\text{eff}}_{7} (\mu_b) |^2  \nonumber \\  &+& \frac{\delta^{NP}_{c}}{m^2_c} {\rm Re} \Bigg[  [\text{C}^{0,\text{eff}}_{7} (\mu_b)]^* \times \bigg ( \text{C}^{0,\text{eff}}_{2} (\mu_b) - \frac{1}{6} \text{C}^{0,\text{eff}}_{1} (\mu_b)\bigg) \Bigg] \Bigg\},\\
{\rm BR}(\bar{B} \to X_s \gamma) &=& \frac{\Gamma (\bar{B} \to X_s \gamma)}{\Gamma_{SL}} {\rm BR}_{SL}.  \label{eq:bsg_final}
\eea 
$ |\bar{D}|$ ($b \to s \gamma$) term, $A$ ($b \to s \gamma g$) term, and the semileptonic width $\Gamma_{SL}$ terms in above expression are implemented from Ref.~\cite{Borzumati1998}.

Our BR$(\bar{B} \to X_s \gamma)$ implementation yields a central  SM prediction value to be $3.39 \times 10^{-4}$, which is extremely close to the SM prediction given in Eq.~(\ref{eq:SMbsg}). From Fig.~\ref{Fig:edmlow} to Fig.~\ref{fig:theta219}, we use coloured bands to show the allowed range of $\pm 2\sigma$ around the experimental central value. We have taken the quadrature of the combined experimental and theoretical uncertainties.  In particular, the grey band is allowed BR$(\bar{B} \to X_s \gamma)$ in the range ($3.32$--$3.77) \times 10^{-4}$ and the green band is allowed range ($2.87$--$3.32) \times 10^{-4}$.\footnote{By coincidence, these limit ranges are same as the $3\sigma$ allowed experimental uncertainty only range.}

\section{Charged Higgs Theoretical and Experimental Constraints}\label{appendix:b}

\subsection{Collider Search Constraints}
The charged Higgs boson production mechanism in a hadron collider can be separated from two cases. 
The first case is when $M_{H^\pm_i}<m_t$ and the dominant production channel is via $t$-quark production and top decay. The second is when  $M_{H^\pm_i}>m_t$  and the dominant production channel is by Higgs-strahlung off $b$-quarks~\cite{Guchait2002,Assamagan2004}. 
The latter one is less limited since the Higgs-strahlung cross section is much smaller than the lighter $t\bar{t}$ production cross section. However, the $t\to b H^+$ signal is overwhelmed by the $t\to bW^+$ background when $M_{H^\pm_i}\approx M_{W^\pm}\approx 80$~GeV. 
Thus, this mass region is still allowed for a charged Higgs state in the 3HDM no matter the decay mode, even at the current Large Hadron Collider (LHC)~\cite{Akeroyd:2018axd,Akeroyd:2019mvt}. 
ATLAS and CMS have performed the searches for a charged scalars via their decay to $\tau\nu$ \cite{CMS:2019bfg}, $cb$ \cite{CMS:2018dzl}, and $cs$ \cite{ATLAS:2013uxj}.
We consider these as constraints on our parameter region, being $\tau\nu$ the most difficult to satisfy.

\subsection{Perturbativity Constraints}
For $\tan\beta$ and $\tan\gamma$, the constraints come from requiring perturbative Yukawa couplings. In here, we follow the approach used for the 2HDM in Ref.~\cite{Barger1990}, in where the condition is that $H^{+} \to t \bar{b}$ computed above the kinematic threshold has to be smaller than half of the mass of charged Higgs ($M_{H^+}/2$).  For example, the lower bound of $\tan\beta$ will come from a Type-I 2HDM constraint~\cite{Barger1990}:

\begin{equation}
	\Gamma(H^+ \to t \bar b) \simeq \frac{3 G_F m_t^2}{4 \sqrt{2} \pi \tan^2\beta} M_{H^+} 
		< \frac{1}{2} M_{H^+}, \qquad {\rm or} \qquad \tan\beta \gtrsim 0.34,
\end{equation}
with $m_t = 173$~GeV.  In the type-II 2HDM, an upper bound on $\tan\beta$ can also be found when the bottom quark Yukawa dominates at large value of the $\tan\beta$, and we have:
\begin{equation}
	\Gamma(H^+ \to t \bar b) \simeq \frac{3 G_F m_b^2 \tan^2\beta}{4 \sqrt{2} \pi} M_{H^+}
		< \frac{1}{2} M_{H^+}, \qquad {\rm or} \qquad \tan\beta \lesssim 125;
\end{equation}
with $m_b \approx 4$~GeV since an even higher upper bound on $\tan\beta$ would be produced using the running bottom quark mass.  
We translate these bounds into the 3HDM, although they are much loser than the collider constraints.
However, having two charged Higgs bosons in 3HDM makes the direct adaptability of the above analysis quite vague.  Instead, we interpret the constraints to bind the Yukawa couplings themselves, in the sense that the previous constraints are equivalent in the 2HDM to $\mathcal{G}_t \lesssim 3.07$ and $\mathcal{G}_b \lesssim 2.90$ in Eq.~(\ref{YukLag}).

In order to match other mixing parameters, we impose the limit $\mathcal{G}_f \lesssim 3$ and derive VEV constraints $v_1 = v \cos\beta \sin\gamma$, $v_2 = v \sin\beta \sin\gamma$ and $v_3 = v \cos\gamma$ using the above quoted $m_t$ and $m_b$ values (with $m_{\tau} = 1.78$~GeV) in the Democratic 3HDM. We find:
\begin{equation}
	\sin\beta \sin\gamma \gtrsim 0.33, \qquad 
	\cos\beta \sin\gamma \gtrsim 0.0077, \qquad
	\tan\gamma \lesssim 290.
\end{equation}
The lower bound on $\tan\gamma$ can be evaluated from the first two constraints,
\begin{equation}
	\tan\gamma \gtrsim 0.35.
\end{equation}
In the condition of $\tan\gamma =1$, the range $0.53 \lesssim \tan\beta \lesssim 92$ is required; the allowed $\tan\beta$ range will be less tighter as $\tan\gamma$ increases. 

\subsection{Top quark width constraints}
Another indirect constraint is coming from the top-quark width ($\Gamma_t$) measurement, which can also limit the charged Higgs boson parameters whenever $M_{H_i^\pm}<m_t$. It can be evaluated by measuring $\Gamma_t$ from the Breit-Wigner resonance reconstruction, which would be produced from the top-quark visible decay products\footnote{Notice that, the single-top cross-section measurement is inapplicable to our analysis because the assumption of top decay channel only covers $t \to bW^+$ without anything else.}.  
The most precise measurement to date is $\Gamma_t=(1.9\pm 0.5)$~GeV~\cite{Zyla2020,ATLAS2019}. In order to satisfy the measurement of $\Gamma_t$, we need to select lower values of $\tan\beta$. 
On the other hand, low $\tan\gamma$ values could let the value of $\Gamma_t$ overpasses the upper bound. 
However, the allowed $\Gamma_t$ can still be satisfied when the two light charged Higgs bosons approach the top-quark mass. 
In this case, very low values of $\tan\gamma$ are also available, and it will be relevant to find a parameter region that survives all the constraints on the top-quark width of collider searches for  $H^\pm_i$ states, EDMs, and BR($\bar{B}\to X_s\gamma$).

\end{paracol}
\reftitle{References}



\bibliographystyle{ieeetr}

%


\end{document}